\documentclass[10pt,fullpage]{article}

\usepackage{fullpage}
\usepackage{times}
\usepackage{amsmath}
\usepackage{amssymb}
\usepackage{amstext}
\usepackage{epsfig}
\usepackage{graphicx}
\usepackage{placeins}
\usepackage{url}
\usepackage{subfig}
\usepackage{paralist}
\usepackage{caption}
\usepackage{multirow}
\usepackage{algorithm}
\usepackage{algorithmic}

\newcommand{\eat}[1]{}

\makeatletter
 {\vskip 5pt\begin{list}{}{%
  \setlength{\topsep}{0pt}\setlength{\partopsep}{0pt}\setlength{\parskip}{0pt}%
  \setlength{\parsep}{0pt}\setlength{\labelwidth}{0pt}%
  \setlength{\rightmargin}{0pt}\setlength{\leftmargin}{0pt}%
  \setlength{\labelsep}{0pt}%
  \obeylines\@vobeyspaces\normalfont\ttfamily%
  \item[]}}
 {\end{list}\vskip5pt\noindent}
\makeatother

\newtheorem{theorem}{Theorem}

\newtheorem{corollary}[theorem]{Corollary}
\newtheorem{definition}{Definition}

\begin{document}

\date{May 2015}

\title{Workload-Driven Vertical Partitioning for Effective Query Processing over Raw Data}

\author{
Weijie Zhao \hspace*{2cm} Yu Cheng \hspace*{2cm} Florin Rusu\\
       \small{University of California, Merced}\\
       \small{5200 N Lake Road}\\
       \small{Merced, CA 95343}\\
       \small\texttt{\{wzhao23,ycheng4,frusu\}@ucmerced.edu}
}

\maketitle

\begin{abstract}

Traditional databases are not equipped with the adequate functionality to handle the volume and variety of ``Big Data''. Strict schema definition and data loading are prerequisites even for the most primitive query session. Raw data processing has been proposed as a schema-on-demand alternative that provides instant access to the data. When loading is an option, it is driven exclusively by the current-running query, resulting in sub-optimal performance across a query workload.
In this paper, we investigate the problem of workload-driven raw data processing with partial loading. We model loading as fully-replicated binary vertical partitioning. We provide a linear mixed integer programming optimization formulation that we prove to be NP-hard. We design a two-stage heuristic that comes within close range of the optimal solution in a fraction of the time. We extend the optimization formulation and the heuristic to pipelined raw data processing, scenario in which data access and extraction are executed concurrently. We provide three case-studies over real data formats that confirm the accuracy of the model when implemented in a state-of-the-art pipelined operator for raw data processing.

\end{abstract}

\section{Introduction}\label{sec:intro}

We are living in the age of ``Big Data'', generally characterized by a series of ``Vs''\footnote{http://www.ibmbigdatahub.com/infographic/four-vs-big-data/}. Data are generated at an unprecedented \textit{volume} by scientific instruments observing the macrocosm~\cite{sdss:original,lsst,ptf:overview} and the microcosm~\cite{lhc:overview,1000genomes:overview}, or by humans connected around-the-clock to mobile platforms such as Facebook and Twitter. These data come in a \textit{variety} of formats, ranging from delimited text to semi-structured JSON and multi-dimensional binaries such as FITS.

The volume and variety of ``Big Data'' pose serious problems to traditional database systems. Before it is even possible to execute queries over a dataset, a relational schema has to be defined and data have to be loaded inside the database. Schema definition imposes a strict structure on the data, which is expected to remain stable. However, this is rarely the case for rapidly evolving datasets represented using key-value and other semi-structured data formats, e.g., JSON. Data loading is a schema-driven process in which data are duplicated in the internal database representation to allow for efficient processing. Even though storage is relatively cheap, generating and storing multiple copies of the same data can easily become a bottleneck for massive datasets. Moreover, it is quite often the case that many of the attributes in the schema are never used in queries.

Motivated by the flexibility of NoSQL systems to access schema-less data and by the Hadoop functionality to directly process data in any format, we have recently witnessed a sustained effort to bring these capabilities inside relational database management systems (RDBMS). Starting with version 9.3, PostgreSQL\footnote{http://www.postgresql.org/} includes support for JSON data type and corresponding functions. Vertica Flex Zone\footnote{http://www.vertica.com/tag/flexzone/} and Sinew~\cite{sinew} implement flex table and column reservoir, respectively, for storing key-value data serialized as maps in a BLOB column. In both systems, certain keys can be promoted to individual columns, in storage as well as in a dynamically evolving schema. With regards to directly processing raw data, several query-driven extensions have been proposed to the loading and external table~\cite{mysql:external-tables,oracle:external-tables} mechanisms. Instead of loading all the columns before querying, in adaptive partial loading~\cite{files-queries-results} data are loaded only at query time, and only the attributes required by the query. This idea is further extended in invisible loading~\cite{invisible-loading}, where only a fragment of the queried columns are loaded, and in NoDB~\cite{nodb}, data vaults~\cite{data-vaults}, SDS/Q~\cite{sdsq}, and RAW~\cite{raw}, where columns are loaded only in memory, but not into the database. SCANRAW~\cite{scanraw} is a super-scalar pipeline operator that loads data speculatively, only when spare I/O resources are available. While these techniques enhance the RDBMS' flexibility to process schema-less raw data, they have several shortcomings, as the following examples show.


\textbf{Example 1: Twitter data.}
The Twitter API\footnote{https://dev.twitter.com/docs/platform-objects/} provides access to several objects in JSON format through a well-defined interface. The schema of the objects is, however, not well-defined, since it includes ``nullable'' attributes and nested objects. The state-of-the-art RDBMS solution to process semi-structured JSON data~\cite{sinew} is to first load the objects as tuples in a BLOB column. Essentially, this entails complete data duplication, even though many of the object attributes are never used. The internal representation consists of a map of key-values that is serialized/deserialized into/from persistent storage. The map can be directly queried from SQL based on the keys, treated as virtual attributes. As an optimization, certain columns -- chosen by the user or by the system based on appearance frequency -- are promoted to physical status. The decision on which columns to materialize is only an heuristic, quite often sub-optimal.

\textbf{Example 2: Sloan Digital Sky Survey (SDSS) data.}
SDSS\footnote{www.sdss.org/dr12/} is a decade-long astronomy project having the goal to build a catalog of all the astrophysical objects in the observable Universe. Images of the sky are taken by a high-resolution telescope, typically in binary FITS format. The catalog data summarize quantities measured from the images for every detected object. The catalog is stored as binary FITS tables. Additionally, the catalog data are loaded into an RDBMS and made available through standard SQL queries. The loading process replicates multi-terabyte data three times -- in ASCII CSV and internal database representation -- and it can take several days---if not weeks~\cite{sdss:sqlLoader}. In order to evaluate the effectiveness of the loading, we extract a workload of 1 million SQL queries executed over the SDSS catalog\footnote{http://skyserver.sdss.org/CasJobs/} in 2014. The most frequent table in the workload is \texttt{photoPrimary}, which appears in more than 70\% of the queries. \texttt{photoPrimary} has 509 attributes, out of which only 74 are referenced in queries. This means that 435 attributes are replicated three times without ever being used---a significantly sub-optimal storage utilization.

\textbf{Problem statement.}
Inspired by the above examples, in this paper, we study the raw data processing with partial loading problem. \textit{Given a dataset in some raw format, a query workload, and a limited database storage budget, find what data to load in the database such that the overall workload execution time is minimized.} This is a standard database optimization problem with bounded constraints, similar to vertical partitioning in physical database design~\cite{raw}. However, while physical design investigates what non-overlapping partitions to build over internal database data, we focus on what data to load, i.e., replicate, in a columnar database with support for multiple storage formats.

Existing solutions for loading and raw data processing are not adequate for our problem. Complete loading not only requires a significant amount of storage and takes a prohibitively long time, but is also unnecessary for many workloads. Pure raw data processing solutions~\cite{nodb,data-vaults,sdsq,raw} are not adequate either, because parsing semi-structured JSON data repeatedly is time-consuming. Moreover, accessing data from the database is clearly optimal in the case of workloads with tens of queries. The drawback of query-driven, adaptive loading methods~\cite{files-queries-results,invisible-loading,scanraw} is that they are greedy, workload-agnostic. Loading is decided based upon each query individually. It is easy to imagine a query order in which the first queries access non-frequent attributes that fill the storage budget, but have limited impact on the overall workload execution time.

\textbf{Contributions.}
To the best of our knowledge, this is the first paper that incorporates query workload in raw data processing. This allows us to model raw data processing with partial loading as fully-replicated binary vertical partitioning. Our contributions are guided by this equivalence. They can be summarized as follows:
\begin{compactitem}
\item We provide a linear mixed integer programming optimization formulation that we prove to be NP-hard and inapproximable.
\item We design a two-stage heuristic that combines the concepts of query coverage and attribute usage frequency. The heuristic comes within close range of the optimal solution in a fraction of the time.
\item We extend the optimization formulation and the heuristic to a restricted type of pipelined raw data processing. In the pipelined scenario, data access and extraction are executed concurrently.
\item We evaluate the performance of the heuristic and the accuracy of the optimization formulation over three real data formats -- CSV, FITS, and JSON -- processed with a state-of-the-art pipelined operator for raw data processing. The results confirm the superior performance of the proposed heuristic over related vertical partitioning algorithms and the accuracy of the formulation in capturing the execution details of a real operator.
\end{compactitem}

\textbf{Outline.}
The paper is organized as follows. Raw data processing, the formal statement of the problem, and an illustrative example are introduced in the preliminaries (Section~\ref{sec:preliminaries}). The mixed integer programming formulation and the proof that the formulation is NP-hard are given in Section~\ref{sec:mip}. The proposed heuristic is presented in detail in Section~\ref{sec:heuristic}. The extension to pipelined raw data processing is discussed in Section~\ref{sec:pipeline}. Extensive experiments that evaluate the heuristic and verify the accuracy of the optimization formulation over three real data formats are presented in Section~\ref{sec:experiments}. Related work on vertical partitioning and raw data processing is briefly discussed in Section~\ref{sec:rel-work}, while Section~\ref{sec:conclusions} concludes the paper.

\section{Preliminaries}\label{sec:preliminaries}

In this section, we introduce query processing over raw data. Then, we provide a formal problem statement and an illustrative example.

\subsection{Query Processing over Raw Data}\label{sec:prelim:in-situ}

\begin{minipage}{.5\textwidth}
Query processing over raw data is depicted in Figure~\ref{fig:scanraw}. The input to the process is a raw file from a non-volatile storage device, e.g., disk or SSD, a schema that can include optional attributes, a procedure to extract tuples with the given schema from the raw file, and a driver query. The output is a tuple representation that can be processed by the query engine and, possibly, is materialized (i.e., loaded) on the same storage device. In the \texttt{READ} stage, data are read from the original raw file, page-by-page, using the file system's functionality. Without additional information about the structure or the content -- stored inside the file or in some external structure -- the entire file has to be read the first time it is accessed. \texttt{EXTRACT} transforms tuples -- one per line -- from raw format into the processing representation, based on the schema provided and using the extraction procedure given as input to the process. There are two stages in \texttt{EXTRACT}---\texttt{TOKENIZE} and \texttt{PARSE}.
\end{minipage}\hfill
\begin{minipage}{.45\textwidth}
	\centering
    \includegraphics[width=\textwidth]{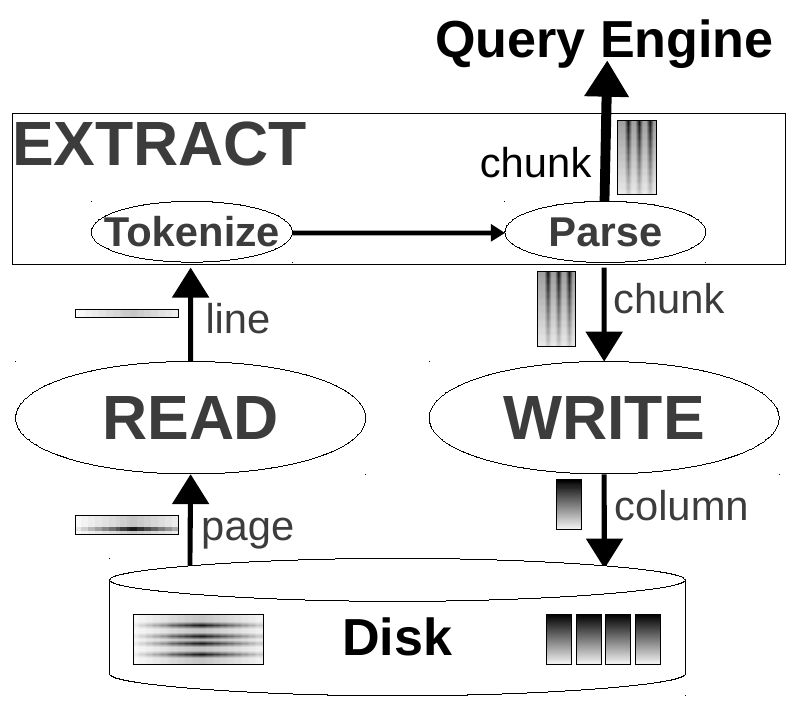}
	\captionof{figure}{Query processing over raw data.}
	\label{fig:scanraw}
\end{minipage}\hfill

\texttt{TOKENIZE} identifies the schema attributes and outputs a vector containing the starting position for every attribute in the tuple---or a subset, if the driver query does not access all the attributes. In \texttt{PARSE}, attributes are converted from raw format to the corresponding binary type and mapped to the processing representation of the tuple---the record in a row-store, or the array in column-stores, respectively. Multiple records or column arrays are grouped into a chunk---the unit of processing. At the end of \texttt{EXTRACT}, data are loaded in memory and ready for query processing. Multiple paths can be taken at this point. In external tables~\cite{mysql:external-tables,oracle:external-tables}, data are passed to the query engine and discarded afterwards. In NoDB~\cite{nodb} and in-memory databases~\cite{instant-loading,data-vaults}, data are kept in memory for subsequent processing. In standard database loading~\cite{files-queries-results,invisible-loading}, data are first written to the database and only then query processing starts. SCANRAW~\cite{scanraw} invokes \texttt{WRITE} concurrently with the query execution, only when spare I/O-bandwidth is available. The interaction between \texttt{READ} and \texttt{WRITE} is carefully scheduled in order to minimize interference.

\subsection{Formal Problem Statement}\label{sec:prelim:problem}

Consider a relational schema $R(A_{1}, A_{2}, \dots, A_{n})$ and an instantiation of it that contains $|R|$ tuples. Semi-structured JSON data can be mapped to the relational model by linearizing nested constructs~\cite{sinew}. In order to execute queries over $R$, tuples have to be read in memory and converted from the storage format into the processing representation. Two timing components correspond to this process. $T_{\textit{RAW}}$ is the time to read data from storage into memory. $T_{\textit{RAW}}$ can be computed straightforwardly for a given schema and storage bandwidth $\textit{band}_{\textit{IO}}$. A constraint specific to raw file processing -- and row-store databases, for that matter -- is that all the attributes are read in a query---even when not required. $T_{\textit{CPU}}$ is the second timing component. It corresponds to the conversion time. For every attribute $A_{j}$ in the schema, the conversion is characterized by two parameters, defined at tuple level. The \textit{tokenizing time $T_{t_{j}}$} is the time to locate the attribute in a tuple in storage format. The \textit{parsing time $T_{p_{j}}$} is the time to convert the attribute from storage format into processing representation. A limited amount of storage $B$ is available for storing data converted into the processing representation. This eliminates the conversion and replaces it with an I/O process that operates at column level---only complete columns can be saved in the processing format. The time to read an attribute $A_{j}$ in processing representation, $T^{\textit{IO}}_{j}$, can be determined when the type of the attribute and $|R|$ are known.

\begin{table}[htbp]
  \begin{center}
    \begin{tabular}{l|cccccccc}

	& $A_{1}$ & $A_{2}$ & $A_{3}$ & $A_{4}$ & $A_{5}$ & $A_{6}$ & $A_{7}$ & $A_{8}$\\

	\hline
	
	$Q_{1}$ & X & X & & & & & & \\

	$Q_{2}$ & X & X & X & X & & & & \\

	$Q_{3}$ & & & X & X & X & & & \\

	$Q_{4}$ & & X & & X & & X & & \\

	$Q_{5}$ & X & & X & X & X & & X & \\

	$Q_{6}$ & X & X & X & X & X & X & X &

    \end{tabular}
  \end{center}

\caption{Query access pattern to raw data attributes.}\label{tbl:prelim:problem:example}
\end{table}

Consider a workload $W = \{Q_{1}, Q_{2}, \dots, Q_{m}\}$ of $m$ SQL-like queries executed over the schema $R$. The workload can be extracted from historical queries or it can be defined by an expert user. Each query $Q_{i}$ is characterized by $\{A_{j_{1}}, A_{j_{2}}, \dots, A_{j_{|Q_{i}|}}\}$, a subset of attributes accessed by the query. Queries are assumed to be distinct, i.e., there are no two queries that access exactly the same set of attributes. A weight $w_{i}$ characterizing importance, e.g., frequency in the workload, is assigned to every query. Ideally, $\sum_{i}{w_{i}}=1$, but this is not necessary.

The problem we investigate in this paper is \textit{how to optimally use the storage $B$ such that the overall query workload execution time is minimized?} Essentially, what attributes to save in processing representation in order to minimize raw file query processing time? We name this problem \textit{raw data processing with partial loading}. We study two versions of the problem---serial and pipeline. In the serial problem, the I/O and the conversion are executed sequentially, while in the pipeline problem, they can overlap. Similar to offline physical database design~\cite{offline-index-building}, the conversion of the attributes stored in processing representation is executed prior to the workload execution. We let the online problem~\cite{online-index-building}, in which conversion and storage are intertwined, for future work.

\subsection{Illustrative Example}\label{sec:prelim:example}

Table~\ref{tbl:prelim:problem:example} depicts the access pattern of a workload of 6 queries to the 8 attributes in a raw file. X corresponds to the attribute being accessed in the respective query. For example, $Q_{1}$ can be represented as $Q_{1} = \{A_{1}, A_{2}\}$. For simplicity, assume that the weights are identical across queries, i.e., $w_{i}=1/6$, $1\leq i \leq 6$. If the amount of storage $B$ that can be used for loading data into the processing representation allows for at most 3 attributes to be loaded, i.e., $B=3$, the problem we address in this paper is what 3 attributes to load such that the workload execution time is minimized? Since $A_{8}$ is not referenced in any of the queries, we are certain that $A_{8}$ is not one of the attributes to be loaded. Finding the best 3 out of the remaining 7 is considerably more difficult.

\section{Mixed Integer Programming}\label{sec:mip}

In order to reason on the complexity of the problem and discuss our solution in a formal framework, we model raw file query processing as mixed integer programming (MIP)~\cite{mip} optimization with 0/1 variables. Table~\ref{tbl:mip:vars} and~\ref{tbl:mip:params} contain the variables and parameters used in the optimization formulation, respectively. Query index $0$ corresponds to saving in the processing representation, i.e., loading, executed before processing the query workload. Parameters include characteristics of the data and the system. The majority of them can be easily determined. The time to tokenize $\textit{T}_{t_{j}}$ and parse $\textit{T}_{p_{j}}$ an attribute are the most problematic since they depend both on the data and the system, respectively. Their value can be configured from previous workload executions or, alternatively, by profiling the execution of the extraction process on a small sample of the raw file.

\begin{table}[htbp]
  \begin{center}
    \begin{tabular}{ll}

	\textbf{Variable} & \textbf{Description}\\

	\hline
	
	$\textit{raw}_{i}$; $i=\overline{0,m}$ & read raw file at query $i$\\
	$t_{ij}$; $i=\overline{0,m}$, $j=\overline{1,n}$ & tokenize attribute $j$ at query $i$\\
	$p_{ij}$; $i=\overline{0,m}$, $j=\overline{1,n}$ & parse attribute $j$ at query $i$\\
	$\textit{read}_{ij}$; $i=\overline{1,m}$, $j=\overline{1,n}$ & read attribute $j$ at query $i$ from processing format\\
	$\textit{save}_{j}$; $j=\overline{1,n}$ & load attribute $j$ in processing format

    \end{tabular}
  \end{center}

\caption{Variables in MIP optimization.}\label{tbl:mip:vars}
\end{table}

The MIP optimization problem for serial raw data processing is formalized as follows (we discuss the pipeline formulation in Section~\ref{sec:pipeline}):
\begin{equation}\label{eq:mip:formulation}
\begin{split}
& \text{minimize} \hspace*{0.2cm} T_{\textit{load}} + \sum_{i=1}^{m}{w_{i}\cdot T_{i}} \hspace*{0.2cm} \text{subject\ to\ constraints:}\\
& C_{1}: \hspace*{0.1cm} \sum_{j=1}^{n}{\textit{save}_{j}} \cdot \textit{SPF}_{j} \cdot |R| \leq B\\
& C_{2}: \hspace*{0.1cm} \textit{read}_{ij} \leq \textit{save}_{j};\ i=\overline{1,m},\ j=\overline{1,n}\\
& C_{3}: \hspace*{0.1cm} \textit{save}_{j} \leq p_{0j} \leq t_{0j} \leq \textit{raw}_{0};\ j=\overline{1,n}\\
& C_{4}: \hspace*{0.1cm} p_{ij} \leq t_{ij} \leq \textit{raw}_{i};\ i=\overline{1,m},\ j=\overline{1,n}\\
& C_{5}: \hspace*{0.1cm} t_{ij} \leq t_{ik};\ i=\overline{0,m},\ j>k=\overline{1,n-1}\\
& C_{6}: \hspace*{0.1cm} \textit{read}_{ij} + p_{ij} = 1;\ i=\overline{1,m},\ j=\overline{1,n},\ A_{j}\in Q_{i}
\end{split}
\end{equation}

\subsection{Objective Function}\label{sec:mip:objective}

The linear objective function consists of two terms. The time to load columns in processing representation $T_{\textit{load}}$ is defined as:
\begin{equation}\label{eq:obj:load}
T_{\textit{load}} = \textit{raw}_{0} \cdot \frac{S_\textit{RAW}}{\textit{band}_{\textit{IO}}} + |R| \cdot \sum_{j=1}^{n}\left({ t_{0j} \cdot T_{t_{j}} + p_{0j} \cdot T_{p_{j}} + \textit{save}_{j} \cdot \frac{\textit{SPF}_{j}}{\textit{band}_{\textit{IO}}} }\right)
\end{equation}
while the execution time corresponding to a query $T_{i}$ is a slight modification:
\begin{equation}\label{eq:obj:query}
T_{i} = \textit{raw}_{i} \cdot \frac{S_\textit{RAW}}{\textit{band}_{\textit{IO}}} + |R| \cdot \sum_{j=1}^{n}\left({ t_{ij} \cdot T_{t_{j}} + p_{ij} \cdot T_{p_{j}} + \textit{read}_{ij} \cdot \frac{\textit{SPF}_{j}}{\textit{band}_{\textit{IO}}} }\right)
\end{equation}
In both cases, the term outside the summation corresponds to reading the raw file. The first term under the sum is for tokenizing, while the second is for parsing. The difference between loading and query execution is only in the third term. In the case of loading, variable $\textit{save}_{j}$ indicates if attribute $j$ is saved in processing representation, while in query execution, variable $\textit{read}_{ij}$ indicates if attribute $j$ is read from the storage corresponding to the processing format at query $i$. We make the reasonable assumption that the read and write I/O bandwidth are identical across storage formats. They are given by $\textit{band}_{\textit{IO}}$.

\subsection{Constraints}\label{sec:mip:constraints}

\begin{minipage}{.4\textwidth}
There are six types of linear constraints in our problem. Constraint $C_{1}$ bounds the amount of storage that can be used for loading data in the processing representation. While $C_{1}$ is a capacity constraint, the remaining constraints are functional, i.e., they dictate the execution of the raw file query processing mechanism. $C_{2}$ enforces that any column read from processing format has to be loaded first. There are $\mathcal{O}(m \cdot n)$ such constraints---one for every attribute in every query. Constraint $C_{3}$ models loading. In order to save a column in processing format, the raw file has to be read and the column has to be tokenized and parsed, respectively.
\end{minipage}\hfill
\begin{minipage}{.6\textwidth}
	\centering
    \begin{tabular}{ll}

	\textbf{Parameter} & \textbf{Description}\\

	\hline
	
	$|R|$ & number of tuples in relation $R$\\
	$S_{\textit{RAW}}$ & size of raw file\\
	$\textit{SPF}_{j}$, $j=\overline{1,n}$ & size of attribute $j$ in processing format\\
	$B$ & size of storage in processing format\\
	$\textit{band}_{\textit{IO}}$ & storage bandwidth\\
	$\textit{T}_{t_{j}}$, $j=\overline{1,n}$ & time to tokenize an instance of attribute $j$\\
	$\textit{T}_{p_{j}}$, $j=\overline{1,n}$ & time to parse an instance of attribute $j$\\
	$w_{i}$, $i=\overline{1,m}$ & weight for query $i$

    \end{tabular}
\captionof{table}{Parameters in MIP optimization.}\label{tbl:mip:params}
\end{minipage}\hfill

While written as a single constraint, $C_{3}$ decomposes into three separate constraints -- one corresponding to each ``$\leq$'' operator -- for a total of $\mathcal{O}(3 \cdot n)$ constraints. $C_{4}$ is a reduced form of $C_{3}$, applicable to query processing. The largest number of constraints, i.e., $\mathcal{O}(m \cdot n^{2})$, in the MIP formulation are of type $C_{5}$. They enforce that it is not possible to tokenize an attribute in a tuple without tokenizing all the preceding schema attributes in the same tuple. $C_{5}$ applies strictly to raw files without direct access to individual attributes. Constraint $C_{6}$ guarantees that every attribute accessed in a query is either extracted from the raw file or read from the processing representation.

\subsection{Computational Complexity}\label{sec:mip:complexity}

There are $\mathcal{O}(m \cdot n)$ binary 0/1 variables in the linear MIP formulation, where $m$ is the number of queries in the workload and $n$ is the number of attributes in the schema. Solving the MIP directly is, thus, impractical for workloads with tens of queries over schemas with hundreds of attributes, unless the number of variables in the search space can be reduced. We prove that this is not possible by providing a reduction from a well-known NP-hard problem to a restricted instance of the MIP formulation. Moreover, we also show that no approximation exists.

\begin{definition}[k-element cover]\label{def:k-elem-cover}
Given a set of $n$ elements $R=\{A_{1}, \dots, A_{n}\}$, $m$ subsets $W=\{Q_{1}, \dots, Q_{m}\}$ of $R$, such that $\bigcup_{i=1}^{m}{Q_{i}}=R$, and a value $k$, the objective in the k-element cover problem is to find a size $k$ subset $R'$ of $R$ that covers the largest number of subsets $Q_{i}$, i.e., $Q_{i}\subseteq R'$, $1\leq i \leq m$.
\end{definition}
For the example in Table~\ref{tbl:prelim:problem:example}, $\{A_{1}, A_{2}\}$ is the single 2-element cover solution (covering $Q_{1}$). While many 3-element cover solutions exist, they all cover only one query.

The k-element cover problem is a restricted instance of the MIP formulation, in which parameters $T_{t_{j}}$, $T_{p_{j}}$, and the loading and reading time to/from database are set to zero, i.e., $\frac{\textit{SPF}_{j}\cdot |R|} {\textit{band}_{\textit{IO}}}\rightarrow 0$, while the raw data reading time is set to one, i.e., $\frac{\textit{S}_{\textit{RAW}}} {\textit{band}_{\textit{IO}}}\rightarrow 1$. The objective function is reduced to counting how many times raw data have to be accessed. The bounding constraint limits the number of attributes that can be loaded, i.e., $\textit{save}_{j}=1$, while the functional constraints determine the value of the other variables. The optimal solution is given by the configuration that minimizes the number of queries accessing raw data. A query does not access raw data when the $\textit{read}_{ij}$ variables corresponding to its attributes are all set to one. When the entire workload is considered, this equates to finding those attributes that cover the largest number of queries, i.e., finding the k-attribute cover of the workload. Given this reduction, it suffices to prove that k-element cover is NP-hard for the MIP formulation to have only exponential-time solutions. We provide a reduction to the well-known minimum k-set coverage problem~\cite{min-k-set-coverage} that proves k-element cover is NP-hard.

\begin{definition}[minimum k-set coverage]\label{def:min-k-set-cover}
Given a set of $n$ elements $R=\{A_{1}, \dots, A_{n}\}$, $m$ subsets $W=\{Q_{1}, \dots, Q_{m}\}$ of $R$, such that $\bigcup_{i=1}^{m}{Q_{i}}=R$, and a value $k$, the objective in the minimum k-set coverage problem is to choose $k$ sets $\{Q_{i_{1}}, \dots Q_{i_{k}}\}$ from $W$ whose union has the smallest cardinality, i.e., $\left|\bigcup_{j=1}^{k}Q_{i_{j}}\right|$.
\end{definition}

\begin{algorithm}
\caption{Reduce $k$-element cover to minimum $k'$-set coverage}\label{alg:reduction-cover-coverage}
\begin{algorithmic}[1]
\REQUIRE Set $R=\{A_{1}, \dots, A_{n}\}$ and $m$ subsets $W=\{Q_{1}, \dots, Q_{m}\}$ of $R$; number $k'$ of sets $Q_{i}$ to choose in minimum set coverage
\ENSURE Minimum number $k$ of elements from $R$ covered by choosing $k'$ subsets from $W$
\FOR{$i$ = 1 to $n$}
	\STATE $\textit{res}$ = \textbf{\textit{k-element cover}}($W$, $i$)
	\STATE \textbf{if} $\textit{res} \geq k'$ \textbf{then} \textbf{return} $i$
\ENDFOR
\end{algorithmic}
\end{algorithm}

Algorithm~\ref{alg:reduction-cover-coverage} gives a reduction from k-element cover to minimum k-set coverage. The solution to minimum k-set coverage is obtained by invoking k-element cover for any number of elements in $R$ and returning the smallest such number for which the solution to k-element cover contains at least $k'$ subsets of $W$. Since we know that minimum k-set coverage is NP-hard~\cite{min-k-set-coverage} and the solution is obtained by solving k-element cover, it implies that k-element cover cannot be any simpler, i.e., k-element cover is also NP-hard. The following theorem formalizes this argument.

\begin{theorem}\label{thm:reduction}
The reduction from k-element cover to minimum k-set coverage given in Algorithm~\ref{alg:reduction-cover-coverage} is correct and complete.
\end{theorem}
\textbf{\textit{Proof.}}
In order to prove the theorem, we have to show that if the answer to the minimum k-set coverage problem is $\textit{ans}$, Algorithm~\ref{alg:reduction-cover-coverage} returns $\textit{ans}$ and if Algorithm~\ref{alg:reduction-cover-coverage} returns $\textit{ans}$, the answer to minimum k-set coverage problem is $\textit{ans}$.

We start with the first implication. Let the optimal solution to the minimum k-set coverage problem be $Q_{\textit{sol}}=\{Q_{i_{1}}, \dots Q_{i_{k}}\}$ and $R_{\textit{sol}}=\{A_{i_{1}}, \dots A_{i_{\textit{ans}}}\}$ be the set of elements in $R$ covered by $Q_{\textit{sol}}$, where $Q_{\textit{sol}} \subseteq W$ and $|R_{\textit{sol}}|=\textit{ans}$. Suppose there exists a subset $R_{\textit{sol}}'$ of $R$ and let the sets covered by $R_{\textit{sol}}'$ be $Q_{\textit{sol}}'$, where $|R_{\textit{sol}}'|<\textit{ans}$ and $|Q_{\textit{sol}}'| \geq k'$. When $|Q_{\textit{sol}}'| \geq k'$, the union of any $k'$ sets in $Q_{\textit{sol}}'$ is no larger than $|R_{\textit{sol}}'|$, which is smaller than $\textit{ans}$. We get a contradiction. Thus, there is no subset of $R$ whose size is smaller than $\textit{ans}$ that covers at least $k'$ sets in $W$. As a result, Algorithm~\ref{alg:reduction-cover-coverage} does not return when $i<\textit{ans}$. By the problem definition, we can use $R_{\textit{sol}}$ to cover at least $k'$ sets $Q_{\textit{sol}}$. Therefore, Algorithm~\ref{alg:reduction-cover-coverage} returns when $i=\textit{ans}$. 

For the second implication, let the elements chosen by the \textit{\textbf{k-element cover}} function be $R_{\textit{sol}}$ and $Q_{\textit{sol}}$ be the sets covered by $R_{\textit{sol}}$, where $|R_{\textit{sol}}|=\textit{ans}$. Suppose the optimal solution to minimum k-set coverage is $Q_{\textit{sol}}'$, which covers elements $R_{\textit{sol}}'$, where $|R_{\textit{sol}}'|<|R_{\textit{sol}}|$ and $|Q_{\textit{sol}}'|=k'$. Then, $Q_{\textit{sol}}'$ is the answer to the function $\textit{\textbf{k-element cover}}(W,|R_{\textit{sol}}'|)$. In this case, Algorithm~\ref{alg:reduction-cover-coverage} returns $|R_{\textit{sol}}'|$ before $\textit{ans}$. We have a contradiction. Therefore, we know that the optimal solution to minimum k-set coverage cannot be smaller than $|R_{\textit{sol}}|$. We can choose any $k'$ sets in $Q_{\textit{sol}}$ as the solution to the minimum k-set coverage problem. The union of the sets in $Q_{\textit{sol}}$ is not larger than $|R_{\textit{sol}}|$. Therefore, the second implication holds.

Based on these two implications, we conclude that the reduction is correct. The fact that Algorithm~\ref{alg:reduction-cover-coverage} has linear time complexity $\mathcal{O}(n)$ guarantees the completeness of the reduction. $\square$

\begin{corollary}\label{thm:reduction:corolar}
The MIP formulation is NP-hard and cannot be approximated unless NP-complete problems can be solved in randomized sub-exponential time.
\end{corollary}
The NP-hardness is a direct consequence of the reduction to the k-element cover problem and Theorem~\ref{thm:reduction}. In addition,~\cite{minzheng} and~\cite{christoph} prove that minimum k-set coverage cannot be approximated within an absolute error of $\frac{1}{2}m^{1-2\epsilon} + \mathcal{O}(m^{1-3\epsilon})$, for any $0 < \epsilon < \frac{1}{3}$, unless P = NP. Consequently, the MIP formulation cannot be approximated.

\section{Heuristic Algorithm}\label{sec:heuristic}

In this section, we propose a novel heuristic algorithm for raw data processing with partial loading that has as a starting point a greedy solution for the k-element cover problem. The algorithm also includes elements from vertical partitioning---a connection we establish in the paper. The central idea is to combine \textit{query coverage} with \textit{attribute usage frequency} in order to determine the best attributes to load. At a high level, query coverage aims at reducing the number of queries that require access to the raw data, while usage frequency aims at eliminating the repetitive extraction of the heavily-used attributes. Our algorithm reconciles between these two conflicting criteria by optimally dividing the available loading budget across them, based on the format of the raw data and the query workload. The solution found by the algorithm is guaranteed to be as good as the solution corresponding to each criterion, considered separately.

In the following, we make the connection with vertical partitioning clear. Then, we present separate algorithms based on query coverage and attribute usage frequency. These algorithms are combined into the proposed heuristic algorithm for raw data processing with partial loading. We conclude the section with a detailed comparison between the proposed heuristic and algorithms designed specifically for vertical partitioning.

\subsection{Vertical Partitioning}\label{sec:heuristic:vertical-part}

Vertical partitioning~\cite{vertical-part} of a relational schema $R(A_{1}, \dots, A_{n})$ splits the schema into multiple schemas -- possibly overlapping -- each containing a subset of the columns in $R$. For example, $\{R_{1}(A_{1}); R_{2}(A_{2}); \dots R_{n}(A_{n})\}$ is the atomic non-overlapping vertical partitioning of $R$ in which each column is associated with a separate partition. Tuple integrity can be maintained either by sorting all the partitions in the same order, i.e., positional equivalence, or by pre-pending a tuple identifier ($\textit{tid}$) column to every partition. Vertical partitioning reduces the amount of data that have to be accessed by queries that operate on a small subset of columns since only the required columns have to be scanned---when they form a partition. However, tuple reconstruction~\cite{tuple-reconstruction} can become problematic when integrity is enforced through $\textit{tid}$ values because of joins between partitions. This interplay between having partitions that contain only the required columns and access confined to a minimum number of partitions, i.e., a minimum number of joins, is the objective function to minimize in vertical partitioning. The process is always workload-driven.

Raw data processing with partial loading can be mapped to \textit{fully-replicated binary vertical partitioning} as follows. The complete raw data containing all the attributes in schema $R$ represent the \textit{raw partition}. The second partition -- \textit{loaded partition} -- is given by the attributes loaded in processing representation. These are a subset of the attributes in $R$. The storage allocated to the loaded partition is bounded. The asymmetric nature of the two partitions differentiates raw data processing from standard vertical partitioning. The raw partition provides access to all the attributes, at the cost of tokenizing and parsing. The loaded partition provides faster access to a reduced set of attributes. In vertical partitioning, all the partitions are equivalent. While having only two partitions may be regarded as a simplification, all the top-down algorithms we are aware of~\cite{vertical-part,chu:vert-part,microsoft:vertical} apply binary splits recursively in order to find the optimal partitions. The structure of raw data processing with partial loading limits the number of splits to one.

\subsection{Query Coverage}\label{sec:heuristic:query-coverage}

\begin{algorithm}
\caption{Query coverage}\label{alg:query-coverage}
\begin{algorithmic}[1]

\REQUIRE Workload $W = \{Q_{1}, \dots, Q_{m}\}$; storage budget $B$
\ENSURE Set of attributes $\{A_{j_{1}}, \dots, A_{j_{k}}\}$ to be loaded in processing representation

\STATE\label{A2:l1} $attsL = \emptyset$; $coveredQ = \emptyset$
\WHILE{$\sum_{j \in attsL} {SPF_{j}} < B$}\label{A2:l2}
	\STATE\label{A2:l3} $idx = \arg\!\max_{i \not\in coveredQ} \left\{ \frac{cost \left(attsL \right) - cost \left(attsL \cup Q_{i} \right)} {\sum_{j \in \{attsL \cup Q_{i} \setminus attsL\}} {SPF_{j}}} \right\}$
	\STATE\label{A2:l4} \textbf{if} $cost \left(attsL \right) - cost \left(attsL \cup Q_{idx} \right) \leq 0$ \textbf{then} \textbf{break}
	\STATE\label{A2:l5} $coveredQ = coveredQ \cup idx$
	\STATE\label{A2:l6} $attsL = attsL \cup Q_{idx}$
\ENDWHILE
\RETURN\label{A2:l8} $attsL$

\end{algorithmic}
\end{algorithm}

A query that can be processed without accessing the raw data is said to be \textit{covered}. In other words, all the attributes accessed by the query are loaded in processing representation. These are the queries whose attributes are contained in the solution to the k-element cover problem. Intuitively, increasing the number of covered queries results in a reduction to the objective function, i.e., total query workload execution time, since only the required attributes are accessed. Moreover, access to the raw data and conversion are completely eliminated. However, given a limited storage budget, it is computationally infeasible to find the optimal set of attributes to load---the k-element cover problem is NP-hard and cannot be approximated (Corollary~\ref{thm:reduction:corolar}). Thus, heuristic algorithms are required.

We design a standard greedy algorithm for the k-element cover problem that maximizes the number of covered queries within a limited storage budget. The pseudo-code is given in Algorithm~\ref{alg:query-coverage}. The solution $\textit{attsL}$ and the covered queries $\textit{coveredQ}$ are initialized with the empty set in line~\ref{A2:l1}. As long as the storage budget is not exhausted (line~\ref{A2:l2}) and the value of the objective function $\textit{cost}$ decreases (line~\ref{A2:l4}), a query to be covered is selected at each step of the algorithm (line~\ref{A2:l3}). The criterion we use for selection is the reduction in the cost function normalized by the storage budget, i.e., we select the query that provides the largest reduction in cost, while using the smallest storage. This criterion gives preference to queries that access a smaller number of attributes and is consistent with our idea of maximizing the number of covered queries. An alternative selection criterion is to drop the cost function and select the query that requires the least number of attributes to be added to the solution. The algorithm is guaranteed to stop when no storage budget is available or all the queries are covered.

\textbf{Example.}
We illustrate how the \textit{\textbf{Query coverage}} algorithm works on the workload in Table~\ref{tbl:prelim:problem:example}. Without loss of generality, assume that all the attributes have the same size and the time to access raw data is considerably larger than the extraction time and the time to read data from processing representation, respectively. These is a common situation in practice, specific to delimited text file formats, e.g., CSV. Let the storage budget be large enough to load three attributes, i.e., $B=3$. In the first step, only queries $Q_{1}$, $Q_{3}$, and $Q_{4}$ are considered for coverage in line~\ref{A2:l3}, due to the storage constraint. While the same objective function value is obtained for each query, $Q_{1}$ is selected for loading because it provides the largest normalized reduction, i.e., $\frac{T_{\textit{RAW}}}{2}$. The other two queries have a normalized reduction of $\frac{T_{\textit{RAW}}}{3}$, where $T_{\textit{RAW}}$ is the time to read the raw data. In the second step of the algorithm, $\textit{attsL} = \{A_{1}, A_{2}\}$. This also turns to be the last step since no other query can be covered in the given storage budget. Notice that, although $Q_{3}$ and $Q_{4}$ make better use of the budget, the overall objective function value is hardly different, as long as reading raw data is the dominating cost component.

\subsection{Attribute Usage Frequency}\label{sec:heuristic:frequency}

\begin{algorithm}
\caption{Attribute usage frequency}\label{alg:freq}
\begin{algorithmic}[1]

\REQUIRE Workload $W=\{Q_{1}, \dots, Q_{m}\}$ of $R$; storage budget $B$; set of loaded attributes $saved=\{A_{s_{1}}, \dots, A_{s_{k}}\}$
\ENSURE Set of attributes $\{A_{s_{k+1}}, \dots, A_{s_{k+t}}\}$ to be loaded in processing representation

\STATE $attsL = saved$
\WHILE{$\sum_{j \in attsL} {SPF_{j}} < B$}\label{A3:l2}
	\STATE\label{A3:l3} $idx = \arg\!\max_{j \not\in attsL} \left\{ cost \left(attsL \right) - cost \left(attsL \cup A_{j} \right) \right\}$
	\STATE $attsL = attsL \cup idx$
\ENDWHILE
\RETURN $attsL$

\end{algorithmic}
\end{algorithm}

The query coverage strategy operates at query granularity. An attribute is always considered as part of the subset of attributes accessed by the query. It is never considered individually. This is problematic for at least two reasons. First, the storage budget can be under-utilized, since a situation where storage is available but no query can be covered, can appear during execution. Second, a frequently-used attribute or an attribute with a time-consuming extraction may not get loaded if, for example, is part of only long queries. The assumption that accessing raw data is the dominant cost factor does not hold in this case. We address these deficiencies of the query coverage strategy by introducing a simple greedy algorithm that handles attributes individually. As the name implies, the intuition behind the attribute usage frequency algorithm is to load those attributes that appear frequently in queries. The rationale is to eliminate the extraction stages that incur the largest cost in the objective function.

The pseudo-code for the attribute usage frequency strategy is given in Algorithm~\ref{alg:freq}. In addition to the workload and the storage budget, a set of attributes already loaded in the processing representation is passed as argument. At each step (line~\ref{A3:l3}), the attribute that generates the largest decrease in the objective function is loaded. In this case, the algorithm stops only when the entire storage budget is exhausted (line~\ref{A3:l2}).

\textbf{Example.}
We illustrate how the \textit{\textbf{Attribute usage frequency}} algorithm works by continuing the example started in the query coverage section. Recall that only two attributes $\textit{saved} = \{A_{1}, A_{2}\}$ out of a total of three are loaded. $A_{4}$ is chosen as the remaining attribute to be loaded since it appears in five queries, the largest number between unloaded attributes. Given that all the attributes have the same size and there is no cost for tuple reconstruction, $\{A_{1}, A_{2}, A_{4}\}$ is the optimal loading configuration for the example in Table~\ref{tbl:prelim:problem:example}.

\subsection{Putting It All Together}\label{sec:heuristic:algorithm}

\begin{algorithm}
\caption{Heuristic algorithm}\label{alg:combined}
\begin{algorithmic}[1]

\REQUIRE Workload $W = \{Q_{1}, \dots, Q_{m}\}$; storage budget $B$
\ENSURE Set of attributes $\{A_{j_{1}}, \dots, A_{j_{k}}\}$ to be loaded in processing representation

\STATE $obj_{min} = \infty$
\FOR {$i = 0$; $i=i+\delta$; $i \leq B$}\label{A4:l2}
	\STATE\label{A4:l3} $attsL_{q} = \textbf{\textit{Query coverage}}(W,i)$
	\STATE\label{A4:l4} $attsL_{f} = \textbf{\textit{Attribute usage frequency}}(W,\Delta_{q},attsL_{q})$
	\STATE $attsL = attsL_{q} \cup attsL_{f}$
	\STATE $obj = cost(attsL)$
	\IF{$obj < obj_{min}$}
		\STATE $obj_{min} = obj$
		\STATE $attsL_{min} = attsL$
	\ENDIF
\ENDFOR
\RETURN $attsL_{min}$

\end{algorithmic}
\end{algorithm}

The heuristic algorithm for raw data processing with partial loading unifies the query coverage and attribute usage frequency algorithms. The pseudo-code is depicted in Algorithm~\ref{alg:combined}. Given a storage budget $B$, \textit{\textbf{Query coverage}} is invoked first (line~\ref{A4:l3}). \textit{\textbf{Attribute usage frequency}} (line~\ref{A4:l4}) takes as input the result produced by \textit{\textbf{Query coverage}} and the unused budget $\Delta_{q}$. Instead of invoking these algorithms only once, with the given storage budget $B$, we consider a series of allocations. $B$ is divided in $\delta$ increments (line~\ref{A4:l2}). Each algorithm is assigned anywhere from $0$ to $B$ storage, in $\delta$ increments. A solution is computed for each of these configurations. The heuristic algorithm returns the solution with the minimum objective. The increment $\delta$ controls the complexity of the algorithm. Specifically, the smaller $\delta$ is, the larger the number of invocations to the component algorithms. Notice, though, that as long as $\frac{B}{\delta}$ remains constant with respect to $m$ and $n$, the complexity of the heuristic remains $\mathcal{O}(m+n)$.

The rationale for using several budget allocations between query coverage and attribute usage frequency lies in the limited view they take for solving the optimization formulation. Query coverage assumes that the access to the raw data is the most expensive cost component, i.e., processing is I/O-bound, while attribute usage frequency focuses exclusively on the extraction, i.e., processing is CPU-bound. However, the actual processing is heavily-dependent on the format of the data and the characteristics of the system. For example, binary formats, e.g., FITS, do not require extraction, while hierarchical text formats, e.g., JSON, require complex parsing. Moreover, the extraction complexity varies largely across data types. The proposed heuristic algorithm recognizes these impediments and solves many instances of the optimization formulation in order to identify the optimal solution.

\subsection{Comparison with Heuristics for Vertical Partitioning}\label{sec:heuristic:comparison-vertical}

As discussed in Section~\ref{sec:heuristic:vertical-part}, raw data processing with partial loading is a special case of vertical partitioning---binary vertical partitioning with full replication. However, there is a fundamental difference between the problem addressed in this paper and standard vertical partitioning. The amount of storage allocated to partitions is not a parameter in vertical partitioning because it is constant and independent of the layout---all the partitions use the same storage, plus-minus metadata. The bounded storage constraint is what makes raw data processing with partial loading a considerably more complicated problem, to which standard vertical partitioning algorithms are not directly applicable.

A comprehensive comparison of vertical partitioning methods is given in~\cite{vert-part:survey}. With few exceptions~\cite{vertical-part,auto-part}, vertical partitioning algorithms consider only the non-replicated case. When replication is considered, it is only partial replication. The bounded scenario -- limited storage budget for replicated attributes -- is discussed only in~\cite{auto-part}. At a high level, vertical partitioning algorithms can be classified along several axes~\cite{vert-part:survey}. We discuss the two most relevant axes for the proposed heuristic. Based on the direction in which partitions are built, we have top-down and bottom-up algorithms. A top-down algorithm~\cite{vertical-part,chu:vert-part,microsoft:vertical} starts with the complete schema and, at each step, splits it into two partitioned schemas. The process is repeated recursively for each resulting schema. A bottom-up algorithm~\cite{hammer:vertical,data-morphing,auto-part,hyrise,jindal:trojan} starts with a series of schemas, e.g., one for each attribute or one for each subset of attributes accessed in a query, and, at each step, merges a pair of schemas into a new single schema. In both cases, the process stops when the objective function cannot be improved further. A second classification axis is given by the granularity at which the algorithm works. An attribute-level algorithm~\cite{hammer:vertical,vertical-part,data-morphing,auto-part,microsoft:vertical,hyrise,jindal:trojan} considers the attributes independent of the queries in which they appear. The interaction between attributes across queries still plays a significant role, though. A query or transaction-level algorithm~\cite{chu:vert-part} works at query granularity. A partition contains either all or none of the attributes accessed in a query.

Based on the classification of vertical partitioning algorithms, the proposed heuristic qualifies primarily as a top-down query-level attribute-level algorithm. However, the recursion is only one-level deep, with the loaded partition at the bottom. The partitioning process consists of multiple steps, though. At each step, a new partition extracted from the raw data is merged into the loaded partition---similar to a bottom-up algorithm. The query coverage algorithm gives the query granularity characteristic to the proposed heuristic, while attribute usage frequency provides the attribute-level property. Overall, the proposed heuristic combines ideas from several classes of vertical partitioning algorithms, adapting their optimal behavior to raw data processing with partial loading. An experimental comparison with specific algorithms is presented in the experiments (Section~\ref{sec:experiments}) and a discussion on their differences in the related work (Section~\ref{sec:rel-work}).

\section{Pipeline Processing}\label{sec:pipeline}

In this section, we discuss on the feasibility of MIP optimization in the case of pipelined raw data processing with partial loading. We consider a super-scalar pipeline architecture in which raw data access and the extraction stages -- tokenize and parse -- can be executed concurrently by overlapping disk I/O and CPU processing. This architecture is introduced in~\cite{scanraw}, where it is shown that, with a sufficiently large number of threads, raw data processing is an I/O-bound task. Loading and accessing data from the processing representation are not considered as part of the pipeline since they cannot be overlapped with raw data access due to I/O interference. We show that, in general, pipelined raw data processing with partial loading cannot be modeled as a linear MIP. However, we provide a linear formulation for a scenario that is common in practice, e.g., binary FITS and JSON format. In these cases, tokenization is atomic. It is executed for all or none of the attributes. This lets parsing as the single variable in the extraction stage. The MIP formulation cannot be solved efficiently, due to the large number of variables and constraints---much larger than in the sequential formulation. We handle this problem by applying a simple modification to the heuristic introduced in Section~\ref{sec:heuristic} that makes the algorithm feasible for pipelined processing.

\subsection{MIP Formulation}\label{sec:pipeline:mip}

Since raw data access and extraction are executed concurrently, the objective function corresponding to pipelined query processing has to include only the maximum of the two:
\begin{equation}\label{eq:obj:pipeline:max}
T^{\textit{pipe}}_{i} = |R| \cdot \sum_{j=1}^{n} \textit{read}_{ij} \cdot \frac{\textit{SPF}_{j}}{\textit{band}_{\textit{IO}}} + \max{\left\{\textit{raw}_{i} \cdot \frac{S_\textit{RAW}}{\textit{band}_{\textit{IO}}}, |R| \cdot \sum_{j=1}^{n}\left(t_{ij} \cdot T_{t_{j}} + p_{ij} \cdot T_{p_{j}}\right) \right\}}
\end{equation}
This is the only modification to the MIP formulation for sequential processing given in Section~\ref{sec:mip}. Since the $\max$ function is non-linear, solving the modified formulation becomes impossible with standard MIP solvers, e.g., CPLEX\footnote{http://www-01.ibm.com/software/commerce/optimization/cplex-optimizer/}, which work only for linear problems. The only alternative is to eliminate the $\max$ function and linearize the objective. However, this cannot be achieved in the general case. It can be achieved, though, for specific types of raw data---binary formats that do not require tokenization, e.g., FITS, and text formats that require complete tuple tokenization, e.g., JSON. As discussed in the introduction, these formats are used extensively in practice.

Queries over raw data can be classified into two categories based on the pipelined objective function in Eq.~(\ref{eq:obj:pipeline:max}). In I/O-bound queries, the time to access raw data is the dominant factor, i.e., $\max$ returns the first argument. In CPU-bound queries, the extraction time dominates, i.e., $\max$ returns the second argument. If the category of the query is known, $\max$ can be immediately replaced with the correct argument and the MIP formulation becomes linear. Our approach is to incorporate the category of the query in the optimization as 0/1 variables. For each query $i$, there is a variable for CPU-bound ($\textit{cpu}_{i}$) and one for IO-bound ($\textit{io}_{i}$). Only one of them can take value 1. Moreover, these variables have to be paired with the variables for raw data access and extraction, respectively. Variables of the form $\textit{cpu}.\textit{raw}_{i}$, $\textit{cpu}.t_{ij}$, and $\textit{cpu}.p_{ij}$ correspond to the variables in Table~\ref{tbl:mip:vars}, in the case of a CPU-bound query. Variables $\textit{io}.\textit{raw}_{i}$, $\textit{io}.t_{ij}$, and $\textit{io}.p_{ij}$ are for the IO-bound case, respectively.

With these variables, we can define the functional \textbf{constraints} for pipelined raw data processing:
\begin{equation}\label{eq:mip:pipeline:func-constraints}
\begin{split}
C_{7}: &\hspace*{0.1cm} \textit{cpu}_{i} + \textit{io}_{i} = 1;\ i=\overline{1,m}\\
C_{8-10}: &\hspace*{0.1cm} \textit{cpu}.x + \textit{io}.x = x;\ x \in \{ \textit{raw}_{i}, t_{ij}, p_{ij}\}\\
C_{11-13}: &\hspace*{0.1cm} \textit{cpu}.x \leq \textit{cpu}_{i};\ i=\overline{1,m}\\
C_{14-16}: &\hspace*{0.1cm} \textit{io}.x \leq \textit{io}_{i};\ i=\overline{1,m}\\
\end{split}
\end{equation}
Constraint $C_{7}$ forces a query to be either CPU-bound or IO-bound. Constraints $C_{8-10}$ tie the new family of CPU/IO variables to their originals in the serial formulation. For example, the raw data is accessed in a CPU/IO query only if it is accessed in the stand-alone query. The same holds for tokenizing/parsing a column $j$ in query $i$. Constraints $C_{11-13}$ and $C_{14-16}$, respectively, tie the value of the CPU/IO variables to the value of the corresponding query variable. For example, only when a query $i$ is CPU-bound, it makes sense for $\textit{cpu}.t_{ij}$ and $\textit{cpu}.p_{ij}$ to be allowed to take value 1. If the query is IO-bound, $\textit{io}.t_{ij}$ and $\textit{io}.p_{ij}$ can be set, but not $\textit{cpu}.t_{ij}$ and $\textit{cpu}.p_{ij}$.

At this point, we have still not defined when a query is CPU-bound and when is IO-bound. This depends on the relationship between the time to access the raw data and the time to extract the referenced attributes. While the parsing time is completely determined by the attributes accessed in the query, the tokenizing time is problematic since it depends not only on the attributes, but also on their position in the schema. For example, in the SDSS \texttt{photoPrimary} table containing 509 attributes, the time to tokenize the $\text{5}^{\text{th}}$ attribute is considerably smaller than the time to tokenize the $\text{205}^{\text{th}}$ attribute. Moreover, there is no linear relationship between the position in the schema and the tokenize time. For this reason, we cannot distinguish between CPU- and IO-bound queries in the general case. However, if there is no tokenization -- the case for binary formats such as FITS -- or the tokenization involves all the attributes in the schema -- the case for hierarchical JSON format -- we can define a threshold $\textit{PT} = \left\lceil \frac {\frac{S_\textit{RAW}}{\textit{band}_{\textit{IO}}} - |R| \cdot \sum_{j=1}^{n} T_{t_{j}}} {\frac {|R| \cdot \sum_{j=1}^{n} T_{p_{j}}} {n}} \right\rceil$ that allows us to classify queries. $\textit{PT}$ is given by the ratio between the time to access raw data less the constant tokenize time and the average time to parse an attribute. Intuitively, $\textit{PT}$ gives the number of attributes that can be parsed in the time required to access the raw data. If a query has to parse more than $\textit{PT}$ attributes, it is CPU-bound. Otherwise, it is IO-bound. The threshold \textbf{constraints} $C_{17}$ and $C_{18}$ make these definitions formal:
\begin{equation}\label{eq:mip:pipeline:threshold-constraints}
\begin{split}
C_{17}: &\hspace*{0.1cm} \sum_{j=1}^{n}{p_{ij}} - \textit{PT} < \textit{cpu}_{i} \cdot n;\ i=\overline{1,m}\\
C_{18}: &\hspace*{0.1cm} \textit{PT} - \sum_{j=1}^{n}{p_{ij}} \leq \textit{io}_{i} \cdot n;\ i=\overline{1,m}\\
\end{split}
\end{equation}
For the atomic tokenization to hold, constraint $C_{5}$ in the serial formulation has to be replaced with $t_{ij} = t_{ik};\ i=\overline{1,m},\ j,k=\overline{1,n-1}$.

The complete pipelined MIP includes the constraints in the serial formulation (Eq.~(\ref{eq:mip:formulation})) and the constraints $C_{7-18}$. The linear \textbf{objective function} corresponding to query processing is re-written using the newly introduced variables as follows:
\begin{equation}\label{eq:obj:pipeline:linear}
T_{i} = \textit{io.raw}_{i} \cdot \frac{S_\textit{RAW}}{\textit{band}_{\textit{IO}}} + |R| \cdot \sum_{j=1}^{n} {\textit{read}_{ij} \cdot \frac{\textit{SPF}_{j}}{\textit{band}_{\textit{IO}}} } + |R| \cdot \sum_{j=1}^{n} {\left( \textit{cpu.t}_{ij} \cdot T_{t_{j}} + \textit{cpu.p}_{ij} \cdot T_{p_{j}} \right)}
\end{equation}

\subsection{Heuristic Algorithm}\label{sec:pipeline:heuristic}

Since the number of variables and constraints increases with respect to the serial MIP formulation, the task of a direct linear solver becomes even harder. It is also important to notice that the problem remains NP-hard and cannot be approximated since the reduction to the k-element cover still applies. In these conditions, heuristic algorithms are the only solution. We design a simple modification to the heuristic introduced in Section~\ref{sec:heuristic} specifically targeted at pipelined raw data processing.

Given a configuration of attributes loaded in processing representation, the category of a query can be determined by evaluating the objective function. What is more important, though, is that the evolution of the query can be traced precisely as attributes get loaded. An I/O-bound query remains I/O-bound as long as not all of its corresponding attributes are loaded. At that point, it is not considered by the heuristic anymore. A CPU-bound query has the potential to become I/O-bound if the attributes that dominate the extraction get loaded. Once I/O-bound, a query cannot reverse to the CPU-bound state. Thus, the only transitions a query can make are from CPU-bound to I/O-bound, and to loaded from there. If an IO-bound query is not covered in the \textit{\textbf{Query coverage}} section of the heuristic, its contribution to the objective function cannot be improved since it cannot be completely covered by \textit{\textbf{Attribute usage frequency}}. Based on this observation, the only strategy to reduce the cost is to select attributes that appear in CPU-bound queries. We enforce this by limiting the selection of the attributes considered in line~\ref{A3:l3} of \textit{\textbf{Attribute usage frequency}} to those attributes that appear in at least one CPU-bound query.

\begin{figure*}[htbp]
\begin{center}
\subfloat[]{\includegraphics[width=0.5\textwidth]{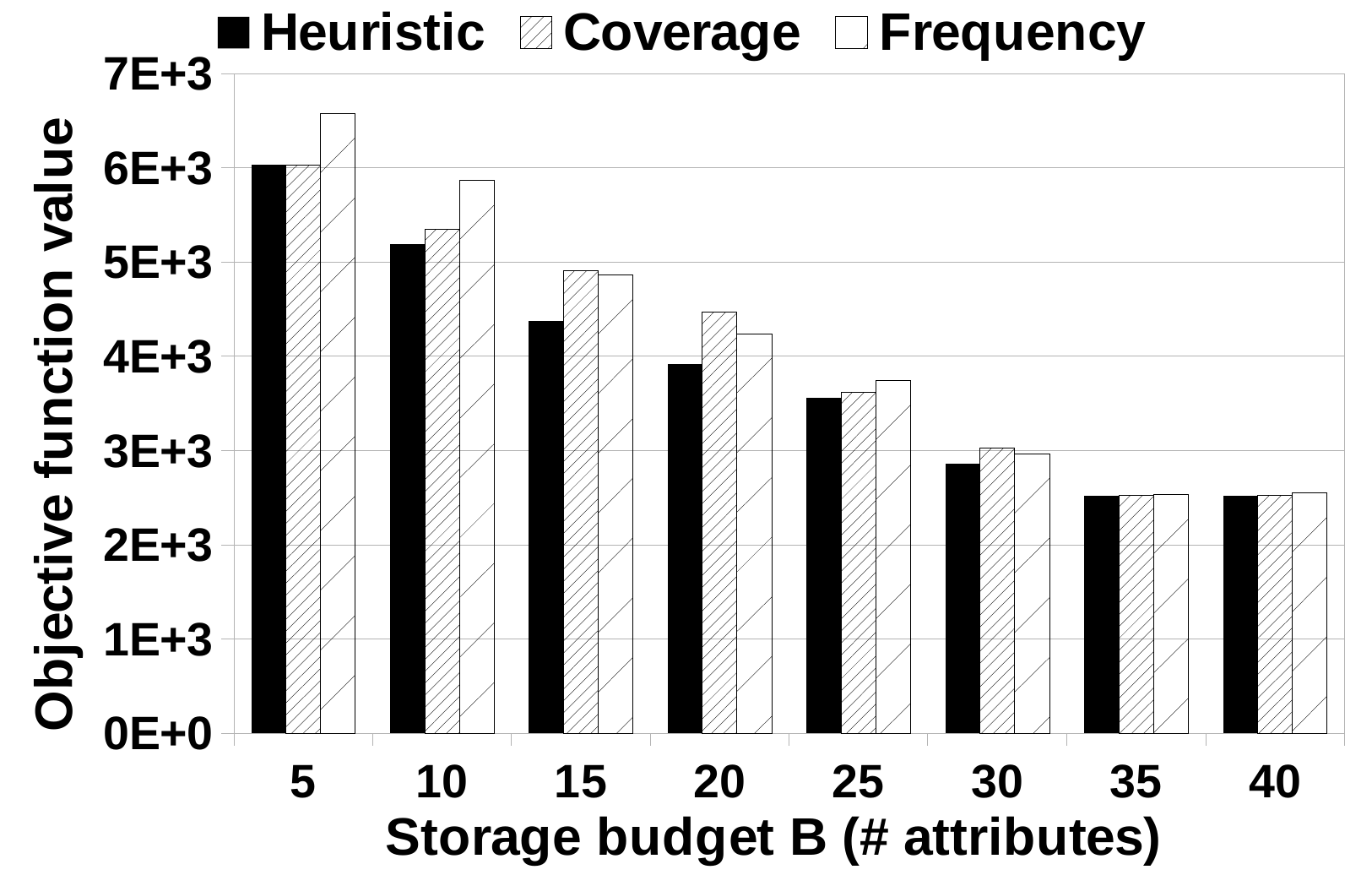}\label{fig:combined_compare}}
\subfloat[]{\includegraphics[width=0.5\textwidth]{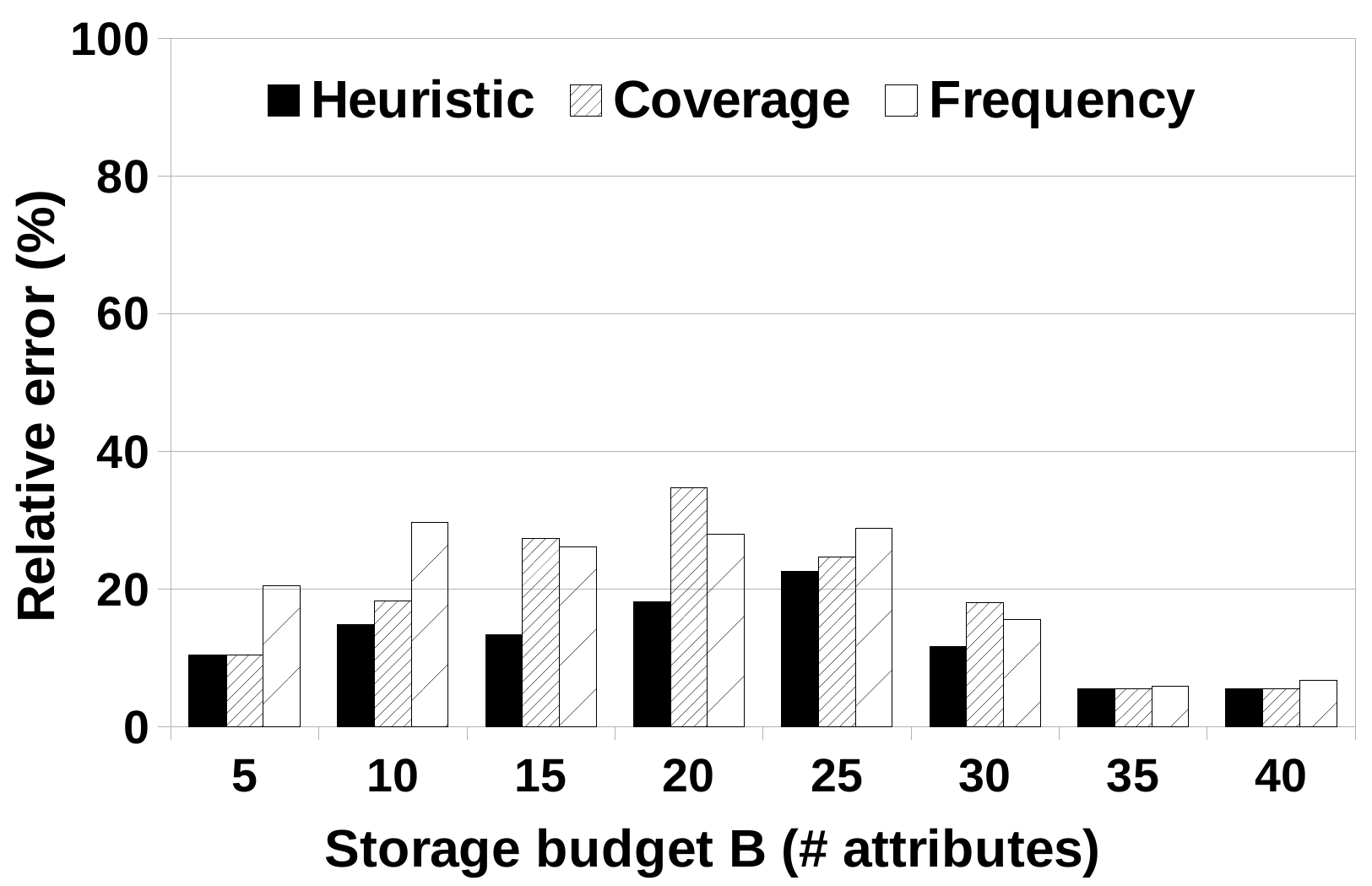}\label{fig:error}}
\caption{Comparison between the stages of the heuristic algorithm: (a) objective function value; (b) relative error with respect to the optimal solution.}
\end{center}
\end{figure*}

\section{Experimental Evaluation}\label{sec:experiments}

The objective of the experimental evaluation is to investigate the accuracy and performance of the proposed heuristic across a variety of datasets and workloads executed sequentially and pipelined. To this end, we explore the accuracy of predicting the execution time for complex workloads over three raw data formats---CSV, FITS, and JSON. Additionally, the sensitivity of the heuristic is quantified with respect to the various configuration parameters. Specifically, the experiments we design are targeted to answer the following questions:

\begin{compactitem}
\item What is the impact of each stage in the overall behavior of the heuristic?
\item How accurate is the heuristic with respect to the optimal solution? With respect to vertical partitioning algorithms?
\item How much faster is the heuristic compared to directly solving the MIP formulation? Compared to other vertical partitioning algorithms?
\item Can the heuristic exploit pipeline processing in partitioning?
\item Do the MIP model and the heuristic reflect reality across a variety of raw data formats?
\end{compactitem}

\textbf{Implementation.}
We implement the heuristic and all the other algorithms referenced in the paper in \texttt{C++}. We follow the description and the parameter settings given in the original paper as closely as possible. The loading and query execution plans returned by the optimization routine are executed with the SCANRAW~\cite{scanraw} operator for raw data processing. SCANRAW supports serial and pipelined execution. The real results returned by SCANRAW are used as reference. We use IBM CPLEX 12.6.1 to implement and solve the MIP formulations. CPLEX supports parallel processing. The number of threads used in the optimization is determined dynamically at runtime.

\textbf{System.}
We execute the experiments on a standard server with 2 AMD Opteron 6128 series 8-core processors (64 bit) -- 16 cores -- 64 GB of memory, and four 2 TB 7200 RPM SAS hard-drives configured RAID-0 in software. Each processor has 12 MB L3 cache while each core has 128 KB L1 and 512 KB L2 local caches. The storage system supports 240, 436 and 1600 MB/second minimum, average, and maximum read rates, respectively---based on the Ubuntu disk utility. \eat{According to \texttt{hdparm}, }The cached and buffered read rates are 3 GB/second and 565 MB/second, respectively. Ubuntu 14.04.2 SMP $64$-bit with Linux kernel 3.13.0-43 is the operating system.

\textbf{Methodology.}
We perform all experiments at least 3 times and report the average value as the result. We enforce data to be read from disk by cleaning the file system buffers before the execution of every query in the workload. This is necessary in order to maintain the validity of the modeling parameters.

\begin{figure*}[htbp]
\begin{center}
\subfloat[]{\includegraphics[width=0.5\textwidth]{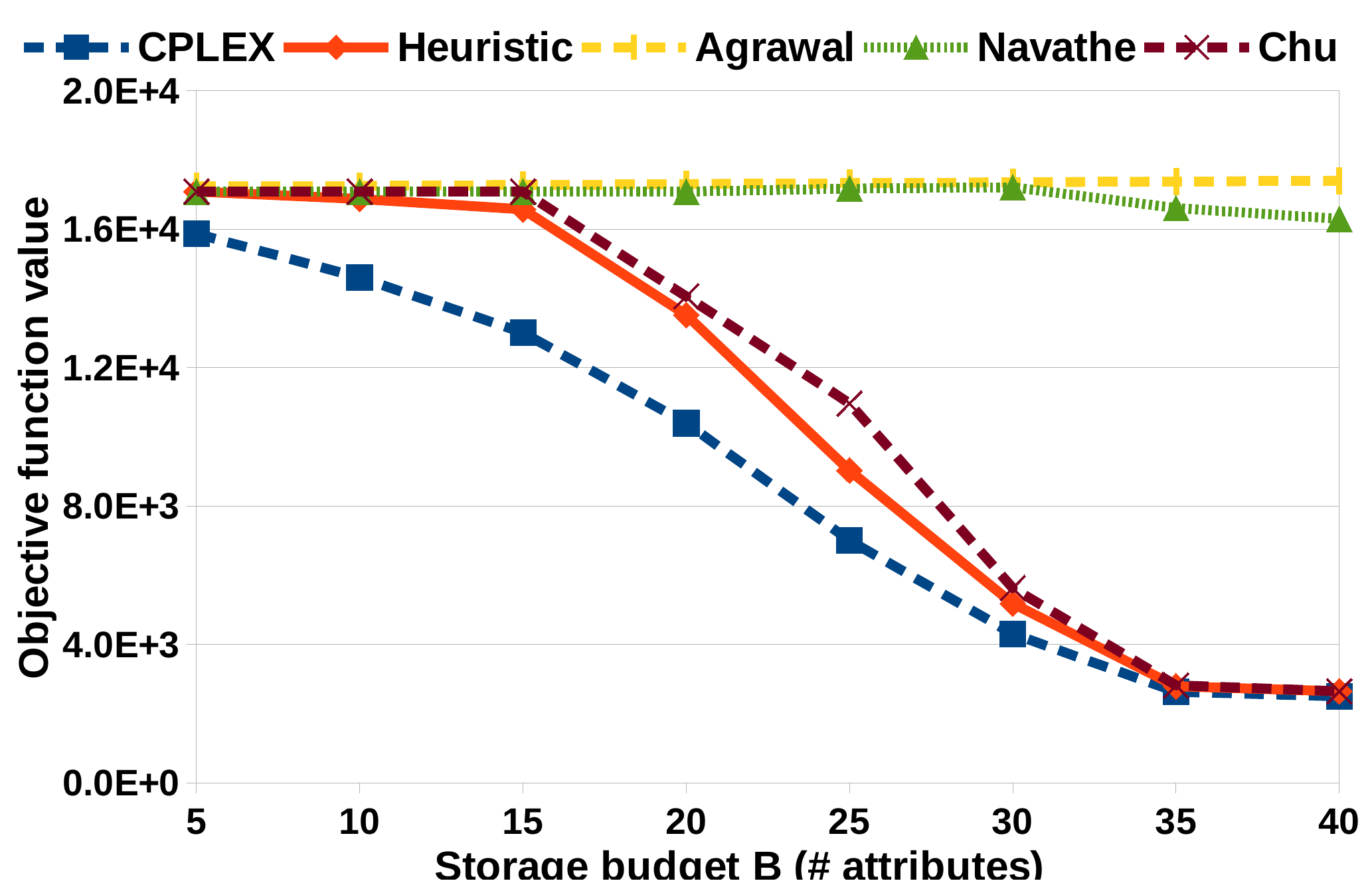}\label{fig:sequential_100_obj}}
\subfloat[]{\includegraphics[width=0.5\textwidth]{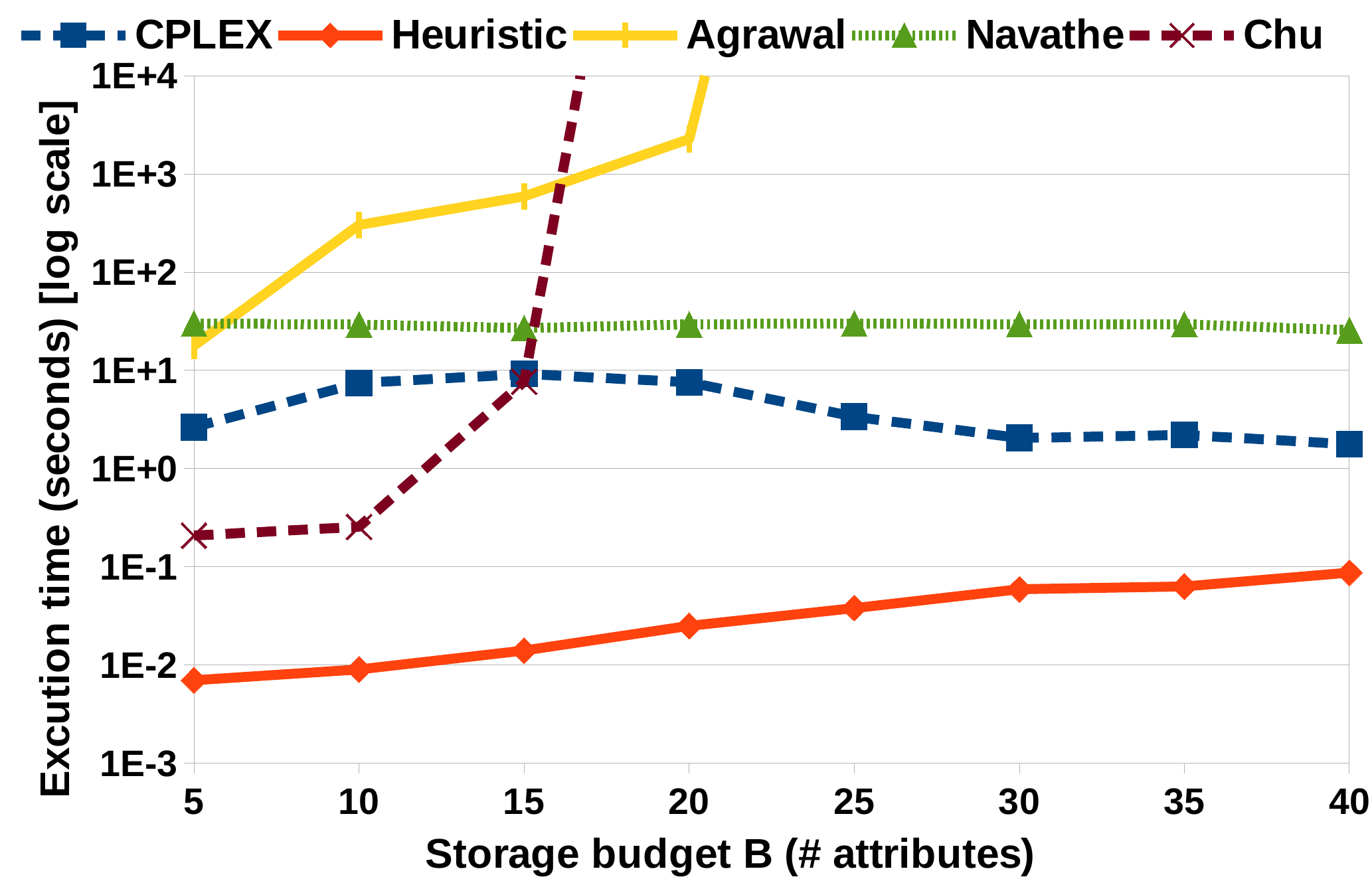}\label{fig:sequential_100_time}}
\caption{Comparison with CPLEX and vertical partitioning algorithms in objective function value (a) and execution time (b) for serial raw data processing.}
\end{center}
\end{figure*}

\textbf{Data.}
We use three types of real data formats in our experiments---CSV, FITS, and JSON. The CSV and FITS data are downloaded from the SDSS project using the CAS tool. They correspond to the complete schema of the \texttt{photoPrimary} table, which contains 509 attributes. The CSV and FITS data are identical. Only their representation is different. CSV is delimited text, while FITS is in binary format. There are 5 million rows in each of these files. CSV is 22 GB in size, while FITS is only 19 GB. JSON is a lightweight semi-structured key-value data format. The Twitter API provides access to user tweets in this format. Tweets have a hierarchical structure that can be flattened into a relational schema. We acquire 5,420,000 tweets by making requests to the Twitter API. There are at most 155 attributes in a tweet. The size of the data is 19 GB.

\textbf{Workloads.}
We extract a real workload of 1 million SQL queries executed over the SDSS catalog in 2014. Out of these, we select the most popular 100 queries over table \texttt{photoPrimary} and their corresponding frequency. These represent approximately 70\% of the 1 million queries. We use these 100 queries as our workload in the experiments over CSV and FITS data. The weight of a query is given by its relative frequency. Furthermore, we extract a subset of the 32 most popular queries and generate a second workload. The maximum number of attributes referenced in both workloads is 74. We create the workload for the tweets data synthetically since we cannot find a real workload that accesses more than a dozen of attributes. The number of attributes in a query is sampled from a normal distribution centered at 20 and having a standard deviation of 20. The attributes in a query are randomly selected out of all the attributes in the schema or, alternatively, out of a subset of the attributes. The smaller the subset, the more attributes are not accessed in any query. The same weight is assigned to all the queries in the workload.

\subsection{Micro-Benchmarks}\label{ssec:experiments:micro-bench}

In this set of experiments, we evaluate the sensitivity of the proposed heuristic with respect to the parameters of the problem, specifically, the number of queries in the workload and the storage budget. We study the impact each stage in the heuristic has on the overall accuracy. We measure the error incurred by the heuristic with respect to the optimal solution computed by CPLEX and the decrease in execution time. We also compare against several top-down vertical partitioning algorithms. We use the SDSS data and workload in our evaluation.

We consider the following vertical partitioning algorithms in our comparison: Agrawal~\cite{microsoft:vertical}, Navathe~\cite{vertical-part}, and Chu~\cite{chu:vert-part}. The \textit{Agrawal algorithm}~\cite{microsoft:vertical} is a pruning-based algorithm in which all the possible column groups are generated based on the attribute co-occurrence in the query workload. For each column group, an interestingness measure is computed. Since there is an exponential number of such column groups, only the ``interesting'' ones are considered as possible partitions. A column group is interesting if the interestingness measure, i.e., CG-Cost, is larger than a specified threshold. The interesting column groups are further ranked based on another measure, i.e., VP-Confidence, which quantifies the frequency with which the entire column group is referenced in queries. The attributes to load are determined by selecting column groups in the order given by VP-Confidence, as long as the storage budget is not filled. While many strategies can be envisioned, our implementation is greedy. It chooses those attributes in a column group that are not already loaded and that minimize the objective function, one-at-a-time. The Agrawal algorithm has exponential complexity $\mathcal{O}\left(2^{n}\right)$ since this is the number of potential column groups. This can be reduced by selecting the CG-Cost threshold intelligently. However, this results in a corresponding accuracy decrease. Trojan Layouts~\cite{jindal:trojan} are a newer version of the Agrawal algorithm in which a different interestingness measure is defined and the selection of the partitions from the interesting column groups is done using an optimal exponential algorithm. Since the authors admit that ``finding the right Trojan Layouts for scientific data sets (having hundreds of attributes), like SDSS, becomes a difficult task to achieve''~\cite{jindal:trojan}, we use the original Agrawal algorithm in our implementation.

The \textit{Navathe algorithm}~\cite{vertical-part} starts with an affinity matrix that quantifies the frequency with which any pair of two attributes appear together in a query. The main step of the algorithm consists in finding a permutation of the rows and columns that groups attributes that co-occur together in queries. This extends upon the affinity of two attributes to a larger number of attributes. While finding the optimal permutation is exponential in the number of attributes, a quadratic greedy algorithm that starts with two random attributes and then chooses the best attribute to add and the best position, one-at-a-time, is given. These are computed based on a benefit function that is independent of the objective. The attributes are ordered on the benefit function in the resulting matrix. The final step of the algorithm consists in finding a split point along the attribute axis that generates two partitions with minimum objective function value across the query workload. An additional condition that we have to consider in our implementation is the storage budget---we find the optimal partition that also fits in the available storage space.

The \textit{Chu algorithm}~\cite{chu:vert-part} considers only those partitions supported by at least one query in the workload, i.e., a column group can be a partition only if it is accessed entirely by a query. Moreover, a column group supported by a query is never split into smaller sub-parts. The algorithm enumerates all the column groups supported by any number of queries in the workload -- from a single query to all the queries -- and chooses the partition that minimizes the objective function. The remaining attributes -- not supported by the query -- form the second partition. This algorithm is exponential in the number of queries in the workload $\mathcal{O}\left(2^{m}\right)$. The solution proposed in~\cite{chu:vert-part} is to limit the number number of query combinations to a relatively small constant, e.g., 5. In our implementation, we let the algorithm run for a limited amount of time, e.g., one hour, and report the best result at that time---if the algorithm has not finished by that time.

\begin{figure*}[htbp]
\begin{center}
\subfloat[]{\includegraphics[width=0.5\textwidth]{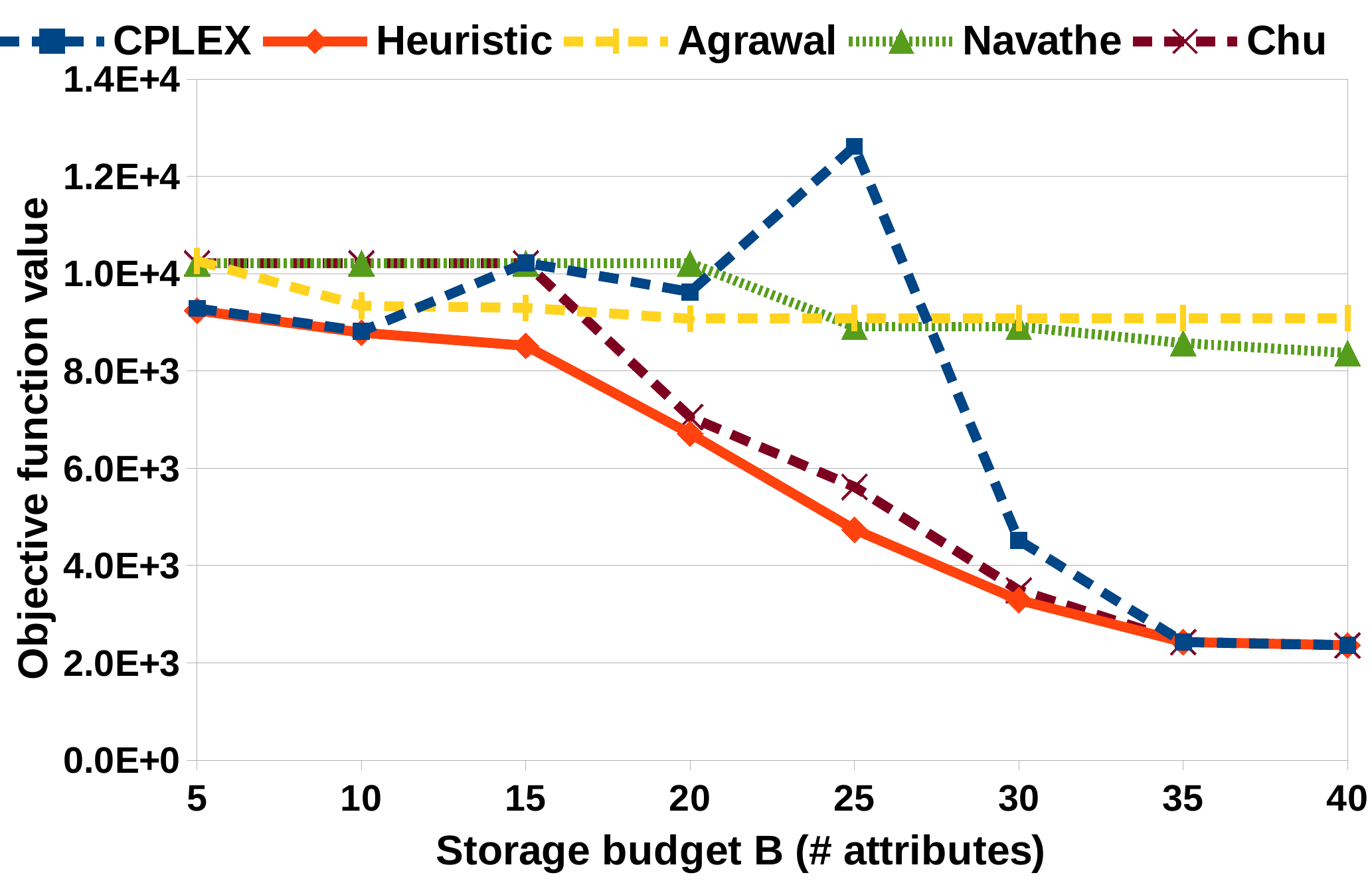}\label{fig:pipeline_100_obj}}
\subfloat[]{\includegraphics[width=0.5\textwidth]{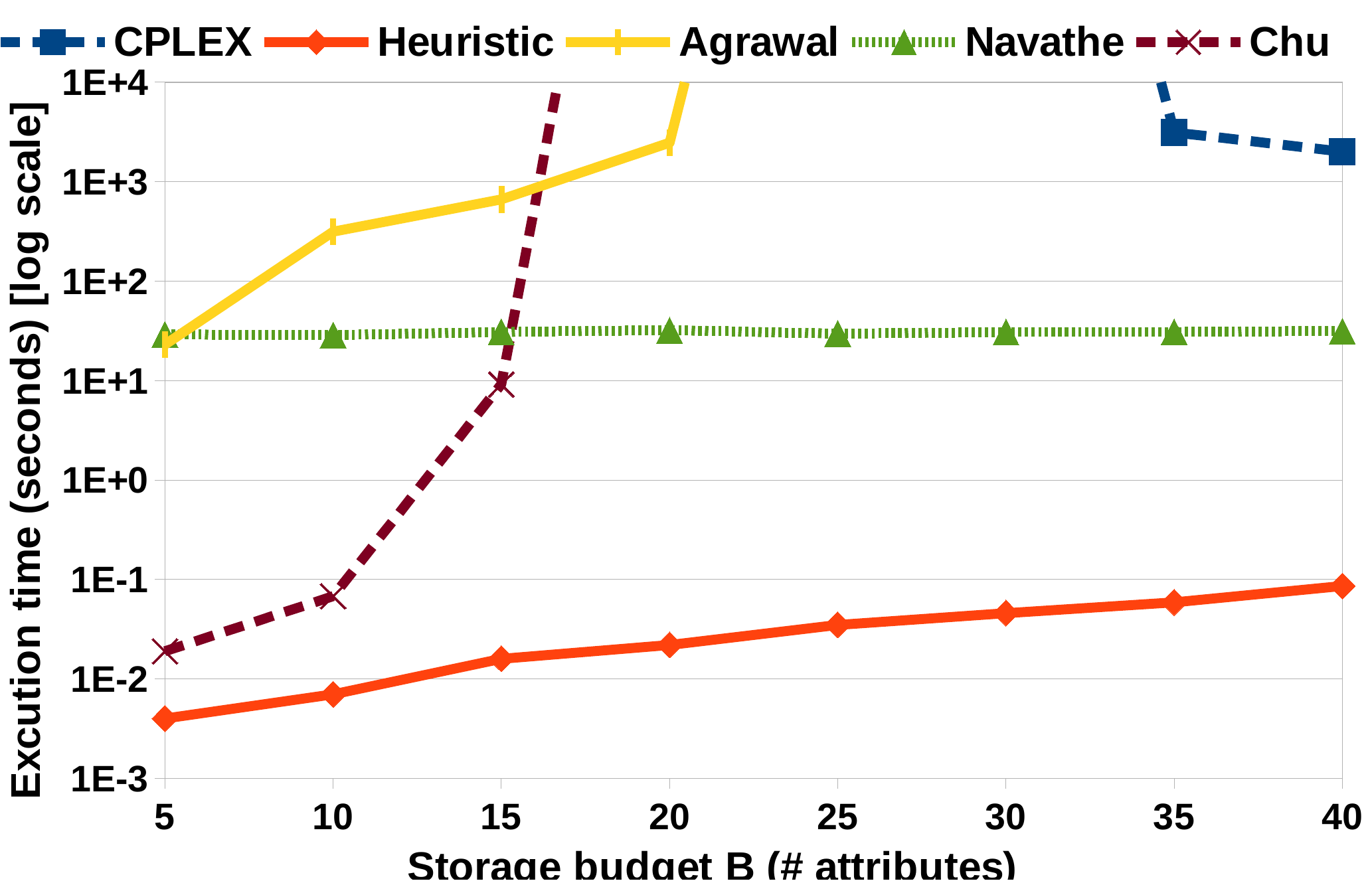}\label{fig:pipeline_100_time}}
\caption{Comparison with CPLEX and vertical partitioning algorithms in objective function value (a) and execution time (b) for pipelined raw data processing.}
\end{center}
\end{figure*}

\textbf{Heuristic stage analysis.}
Figure~\ref{fig:combined_compare} and~\ref{fig:error} depict the impact each stage in the heuristic -- query \textbf{coverage} and attribute usage \textbf{frequency} -- has on the accuracy, when taken separately and together, i.e., \textbf{heuristic}. We measure both the absolute value (Figure~\ref{fig:combined_compare}) and the relative error with respect to the optimal value (Figure~\ref{fig:error}). We depict these values as a function of the storage budget, given as the number of attributes that can be loaded. We use the 32 queries workload. As expected, when the budget increases, the objective decreases. In terms of relative error, though, the heuristic is more accurate at the extremes---small budget or large budget. When the budget is medium, the error is the highest. The reason for this behavior is that, at the extremes, the number of choices for loading is considerably smaller and the heuristic finds a good enough solution. When the storage budget is medium, there are many loading choices and the heuristic makes only local optimal decisions that do not necessarily add-up to a good global solution. The two-stage heuristic has better accuracy than each stage taken separately. This is more clear in the case of the difficult problems with medium budget. Between the two separate stages, none of them is dominating the other in all the cases. This proves that our integrated heuristic is the right choice since it always improves upon the best stage taken individually.

\textbf{Serial heuristic accuracy.}
Figure~\ref{fig:sequential_100_obj} depicts the accuracy as a function of the storage budget for several algorithms in the case of serial raw data processing. The workload composed of 100 queries is used. Out of the heuristic algorithms, the proposed heuristic is the most accurate. As already mentioned, the largest error is incurred when the budget is medium. Between the vertical partitioning algorithms, the query-level granularity algorithm~\cite{chu:vert-part} is the most accurate. The other two algorithms~\cite{vertical-part,microsoft:vertical} do not improve as the storage budget increases. This is because they are attribute-level algorithms that are not optimized for covering queries.

\textbf{Serial heuristic execution time.}
Figure~\ref{fig:sequential_100_time} depicts the execution time for the same scenario as in Figure~\ref{fig:sequential_100_obj}. It is clear that the proposed heuristic is always the fastest, even by three orders of magnitude in the best case. Surprisingly, calculating the exact solution using CPLEX is faster than all the vertical partitioning algorithms almost in all the cases. If an algorithm does not finish after one hour, we stop it and take the best solution at that moment. This is the case for Chu~\cite{chu:vert-part} and Agrawal~\cite{microsoft:vertical}. However, the solution returned by Chu is accurate---a known fact from the original paper.

\textbf{Pipelined heuristic accuracy.}
The objective function value for pipelined processing over FITS data is depicted in Figure~\ref{fig:pipeline_100_obj}. The same 100 query workload is used. The only difference compared to the serial case is that CPLEX cannot find the optimal solution in less than one hour. However, it manages to find a good-enough solution in most cases. The proposed heuristic achieves the best accuracy for all the storage budgets.

\textbf{Pipelined heuristic execution time.}
The proposed heuristic is the only solution that achieves sub-second execution time for all the storage budgets (Figure~\ref{fig:pipeline_100_time}). CPLEX finishes execution in the alloted time only when the budget is large. The number of variables and constraints in the pipeline MIP formulation increase the search space beyond what the CPLEX algorithms can handle.

\subsection{Case Study: CSV Format}\label{ssec:experiments:csv}

We provide a series of case studies over different data formats in order to validate that the raw data processing architecture depicted in Figure~\ref{fig:scanraw} is general and the MIP models corresponding to this architecture fit reality. We use the implementation of the architecture in the SCANRAW operator~\cite{scanraw} as a baseline. For a given workload and loading plan, we measure the cumulative execution time after each query and compare the result with the estimation computed by the MIP formulation. If the two match, this is a good indication that the MIP formulation models reality accurately.
\begin{minipage}{.45\textwidth}
The CSV format maps directly to the raw data processing architecture. In order to apply the MIP formulation, the value of the parameters has to be calibrated for a given system and a given input file. The time to tokenize $T_{t_{j}}$ and parse $T_{p_{j}}$ an attribute are the only parameters that require discussion. This can be done by executing the two stages on a sample of the data and measuring the average value of the parameter for each attribute. As long as accurate estimates are obtained, the model will be accurate. Figure~\ref{fig:csv} confirms this on the SDSS workload of 32 queries. In this case, there is a perfect match between the model and the SCANRAW execution.
\end{minipage}\hfill
\begin{minipage}{.5\textwidth}
	\centering
	\includegraphics[width=\textwidth]{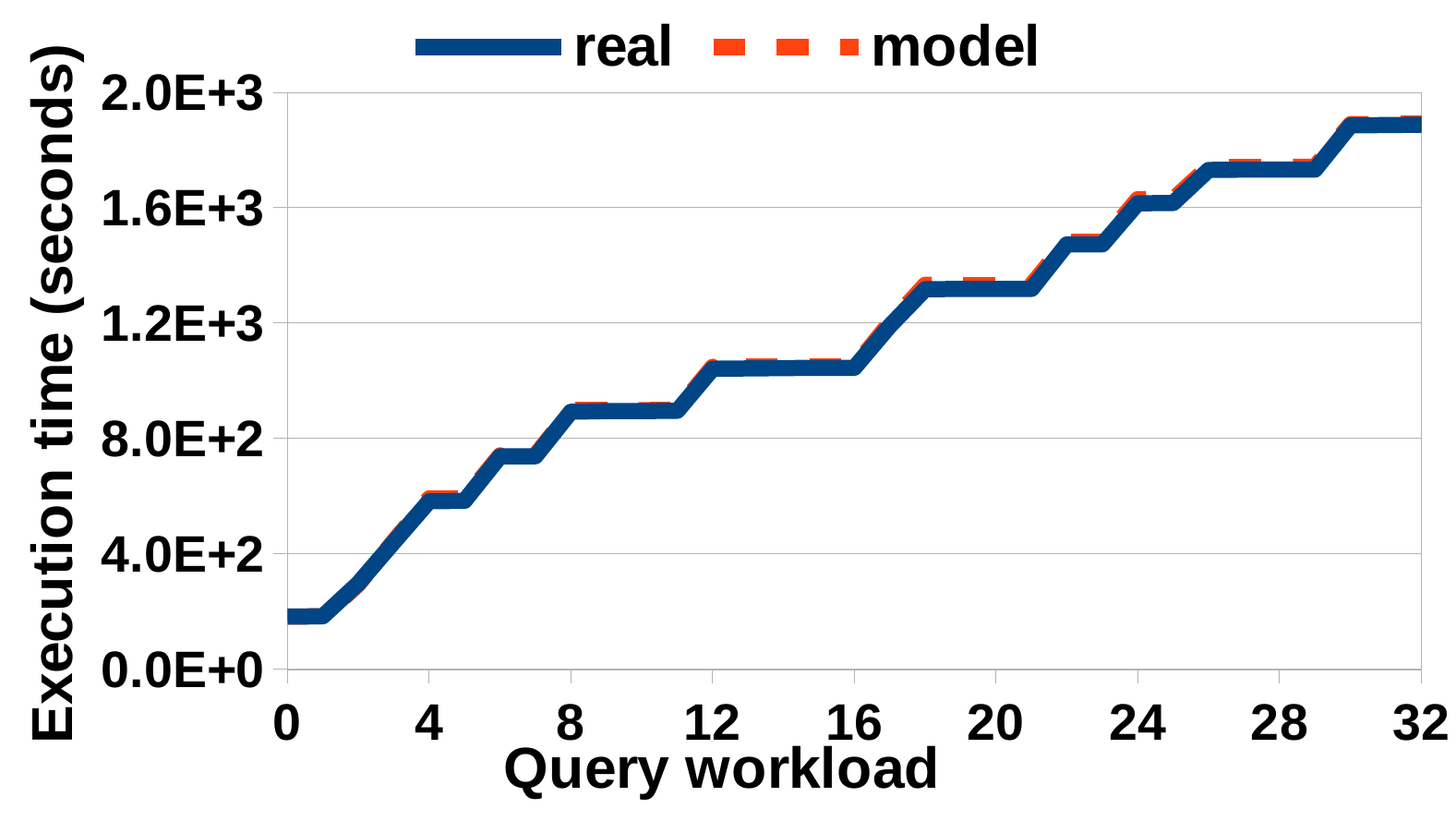}
	\captionof{figure}{Serial CSV model validation.}
	\label{fig:csv}
\end{minipage}\hfill

\subsection{Case Study: FITS Format}\label{ssec:experiments:fits}

\begin{minipage}{.45\textwidth}
Since FITS is a binary format, there is no extraction phase, i.e., tokenizing and parsing, in the architecture. Moreover, data can be read directly in the processing representation, as long as the file access library provides such a functionality. CFITSIO -- the library we use in our implementation -- can read a range of values of an attribute in a pre-allocated memory buffer. However, we observed experimentally that, in order to access any attribute, there is a high startup time. Essentially, the entire data are read in order to extract the attribute. The additional time is linear in the number of attributes.
\end{minipage}\hfill
\begin{minipage}{.5\textwidth}
	\centering
	\includegraphics[width=\textwidth]{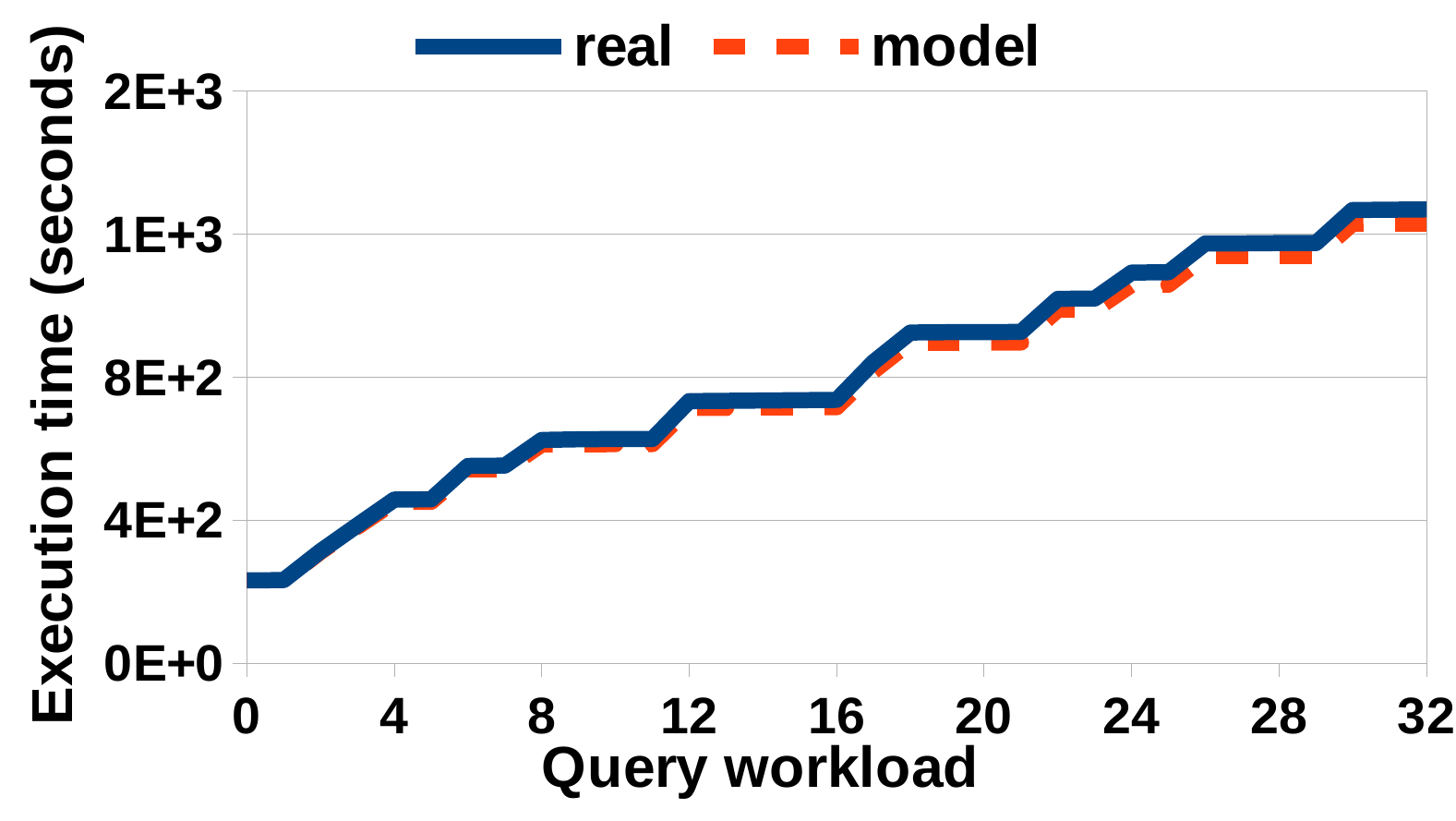}
	\captionof{figure}{Serial FITS model validation.}
	\label{fig:fits}
\end{minipage}\hfill

Based on these observations -- that may be specific to CFITSIO -- the following parameters have to be calibrated: the time to read the raw data corresponds to the startup time; an extraction time proportional with the number of attributes in the query is the equivalent of $T_{p_{j}}$. $T_{t_{j}}$ is set to zero. Although pipelining is an option for FITS data, due to the specifics of the CFITSIO\footnote{http://heasarc.gsfc.nasa.gov/fitsio/fitsio.html} library, it is impossible to apply it. The result for the SDSS data confirms that the model is a good fit for FITS data since there is almost complete overlap in Figure~\ref{fig:fits}.

\subsection{Case Study: JSON Format}\label{ssec:experiments:json}

At first sight, it seems impossible to map JSON data on the raw data processing architecture and the MIP model. Looking deeper, we observe that JSON data processing is even simpler than CSV processing. This is because every object is fully-tokenized and parsed in an internal map data structure, independent of the requested attributes. At least this is how the JSONCPP\footnote{http://sourceforge.net/projects/jsoncpp/} library works.\\
\begin{minipage}{.45\textwidth}
Once the map is built, it can be queried for any key in the schema. For schemas with a reduced number of hierarchical levels -- the case for tweets -- there is no difference in query time across levels. Essentially, the query time is proportional only with the number of requested keys, independent of their existence or not. Based on these observations, we set the model parameters as follows. $T_{t_{j}}$ is set to the average time to build the map divided by the maximum number of attributes in the schema. $T_{p_{j}}$ is set to the map data structure query time. Since $T_{t_{j}}$ is a constant, the pipelined MIP formulation applies to the JSON format. The results in Figure~\ref{fig:json} confirm the accuracy of the model over a workload of 32 queries executed in SCANRAW.
\end{minipage}\hfill
\begin{minipage}{.5\textwidth}
	\centering
	\includegraphics[width=\textwidth]{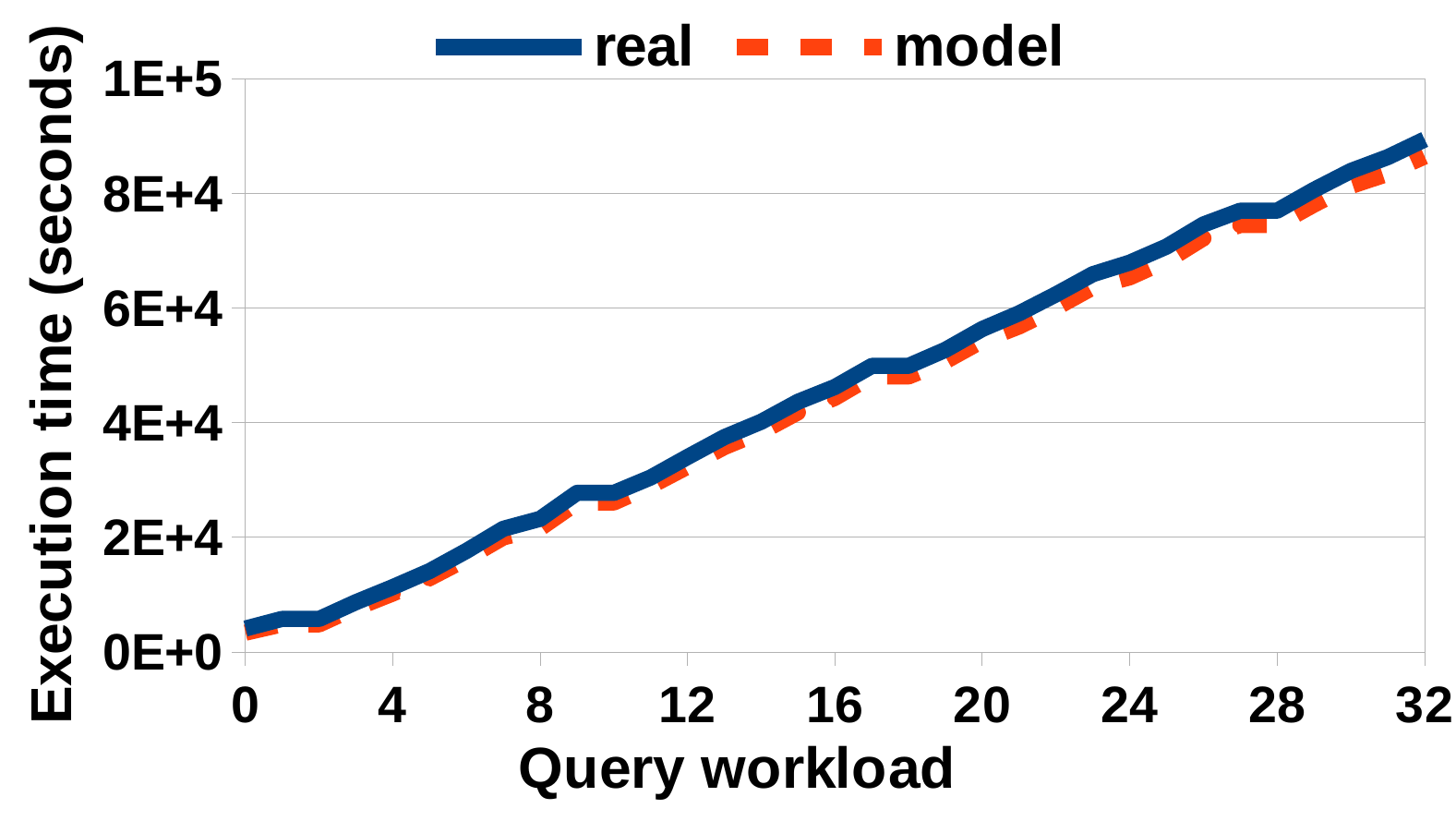}
	\captionof{figure}{Pipelined JSON model validation.}
	\label{fig:json}
\end{minipage}\hfill

\subsection{Discussion}\label{ssec:experiments:discussion}

The experimental evaluation provides answers to each of the questions raised at the beginning of the section. The two-stage heuristic improves over each of the component parts. It is not clear which of the query coverage and attribute usage frequency is more accurate. Using them together guarantees the best results. The proposed heuristic comes close to the optimal solution whenever the storage budget is either small or large. When many choices are available -- the case for a medium budget -- the accuracy decreases, but remains superior to the accuracy of the other vertical partitioning methods. In terms of execution time, the proposed heuristic is the clear winner---by as much as three orders of magnitude. Surprisingly, CPLEX outperforms the other heuristics in the serial case. This is not necessarily unexpected, given that these algorithms have been introduced more than two decades ago. The case studies confirm the applicability of the MIP formulation model to several raw data formats. The MIP model fits the reality almost perfectly both for serial and pipelined raw data processing.

\section{Related Work}\label{sec:rel-work}

Two lines of research are most relevant to the work presented in this paper---raw data processing and vertical partitioning as a physical database design technique. Our contribution is to integrate workload information in raw data processing and model the problem as vertical partitioning optimization. To the best of our knowledge, this is the first paper to consider the problem of optimal vertical partitioning for raw data processing with partial loading.

\textbf{Raw data processing.}
Several methods have been proposed for processing raw data within a database engine. The vast majority of them bring enhancements to the external table functionality, already supported by several major database servers~\cite{oracle:external-tables,mysql:external-tables}. A common factor across many of these methods is that they do not consider loading converted data inside the database. At most, data are cached in memory on a query-by-query basis. This is the approach taken in NoDB~\cite{nodb}, Data Vaults~\cite{data-vaults}, SDS/Q~\cite{sdsq}, RAW~\cite{raw}, and Impala~\cite{impala}. Even when loading is an option, for example in adaptive partial loading~\cite{files-queries-results}, invisible loading~\cite{invisible-loading}, and SCANRAW~\cite{scanraw}, the workload is not taken into account and the storage budget is unlimited. The decision on what to load is local to every query, thus, prone to be acutely sub-optimal over the entire workload. 

The heuristic developed in this paper requires workload knowledge and aims to identify the optimal data to load such that the execution time of the entire workload is minimized. As in standard database processing, loading is executed offline, before query execution. However, the decision on what data to load is intelligent and the time spent on loading is limited by the allocated storage budget. Notice that the heuristic is applicable both to secondary storage-based loading as well as to one-time in-memory caching without subsequent replacement.

\textbf{Vertical partitioning.}
Vertical partitioning has a long-standing history as a physical database design strategy, dating back to the 1970's. Many types of solutions have been proposed over the years, ranging from integer programming formulations to top-down and bottom-up heuristics that operate at the granularity of a query or of an attribute. A comparative analysis of several vertical partitioning algorithms is presented in~\cite{vert-part:survey}. The serial MIP formulation for raw data processing is inspired from the formulations for vertical partitioning given in~\cite{hoffer:mip,psyu:mip}. While both are non-linear, none of these formulations considers pipeline processing. We prove that even the linear MIP formulation is NP-hard. The scale of the previous results for solving MIP optimizations have to be taken with a grain of salt, given the extensive enhancements to integer programming solvers over the past two decades. As explained in Section~\ref{sec:heuristic:comparison-vertical}, the proposed heuristic combines ideas from several classes of vertical partitioning algorithms, adapting their optimal behavior to raw data processing with partial loading. The top-down transaction-level algorithm given in~\cite{chu:vert-part} is the closest to the query coverage stage. While query coverage is a greedy algorithm, \cite{chu:vert-part} employs exhaustive search to find the solution. As the experimental results show, this is time-consuming. Other top-down heuristics~\cite{vertical-part,microsoft:vertical} consider the interaction between attributes across the queries in the workload. The partitioning is guided by a quantitative parameter that measures the strength of the interaction. In~\cite{vertical-part}, only the interaction between pairs of attributes is considered. The attribute usage frequency phase of the proposed heuristic treats each attribute individually, but only after query coverage is executed. The objective in~\cite{microsoft:vertical} is to find a set of vertical partitions that are subsequently evaluated for index creation. Since we select a single partitioning scheme, the process is less time-consuming. Finally, the difference between the proposed heuristic and bottom-up algorithms~\cite{hammer:vertical,data-morphing,auto-part,hyrise,jindal:trojan} is that the latter cannot guarantee that only two partitions are generated at the end. This is a requirement for raw data processing with partial loading. All these algorithms are offline. They are executed only once, before query processing, over a known workload. Online vertical partitioning algorithms form a separate class. In~\cite{online-index-building}, the entire workload is known in advance. However, the order of the queries is fixed and the vertical partitioning evolves. Another series of algorithms~\cite{ailamaki:smdb:online-vert-part,ailamaki:ssdbm:online-vert-part,jindal:o2p} operates over an unknown workload, given one query at a time. Their goal is to gather evidence from the past workload in order to determine the optimal vertical partitioning at each query.

\section{Conclusions and Future Work}\label{sec:conclusions}

In this paper, we study the problem of workload-driven raw data processing with partial loading. We model loading as binary vertical partitioning with full replication. Based on this equivalence, we provide a linear mixed integer programming optimization formulation that we prove to be NP-hard and inapproximable. We design a two-stage heuristic that combines the concepts of query coverage and attribute usage frequency. The heuristic comes within close range of the optimal solution in a fraction of the time. We extend the optimization formulation and the heuristic to a restricted type of pipelined raw data processing. In the pipelined scenario, data access and extraction are executed concurrently. We evaluate the performance of the heuristic and the accuracy of the optimization formulation over three real data formats -- CSV, FITS, and JSON -- processed with a state-of-the-art pipelined operator for raw data processing. The results confirm the superior performance of the proposed heuristic over related vertical partitioning algorithms and the accuracy of the formulation in capturing the execution details of a real operator.

Following the steps of database physical design, we envision several avenues to extend the proposed research in the future. We can move from the offline loading setting to online loading, where query processing and loading are intertwined. We can assume that the workload is known beforehand or it is given one query at a time. We can drop the strict requirement of atomic attribute loading and allow for portions -- horizontal partitions -- of an attribute to be loaded. Finally, we can consider a multi-query processing environment in which raw data access and attribute extraction can be shared across several queries.

\paragraph*{Acknowledgments.}
This work is supported by a U.S. Department of Energy Early Career Award (DOE Career).

\bibliographystyle{abbrv}

\begin{thebibliography}{10}

\bibitem{sdss:original}
{A. Szalay et al.}
\newblock {Designing and Mining Multi-Terabyte Astronomy Archives: The Sloan
  Digital Sky Survey}.
\newblock In {\em {Proceedings of 2000 ACM SIGMOD International Conference on
  Management of Data}}, pages 451--462, 2000.

\bibitem{invisible-loading}
A.~Abouzied, D.~Abadi, and A.~Silberschatz.
\newblock {Invisible Loading: Access-Driven Data Transfer from Raw Files into
  Database Systems}.
\newblock In {\em {Proceedings of 2013 EDBT/ICDT Extended Database Technology
  Conference}}, pages 1--10, 2013.

\bibitem{online-index-building}
S.~Agrawal, E.~Chu, and V.~Narasayya.
\newblock {Automatic Physical Design Tuning: Workload as a Sequence}.
\newblock In {\em {Proceedings of 2006 ACM SIGMOD International Conference on
  Management of Data}}, pages 683--694, 2006.

\bibitem{microsoft:vertical}
S.~Agrawal, V.~Narasayya, and B.~Yang.
\newblock {Integrating Vertical and Horizontal Partitioning into Automated
  Physical Database Design}.
\newblock In {\em {Proceedings of 2004 ACM SIGMOD International Conference on
  Management of Data}}, pages 359--370, 2004.

\bibitem{nodb}
I.~Alagiannis, R.~Borovica, M.~Branco, S.~Idreos, and A.~Ailamaki.
\newblock {NoDB: Efficient Query Execution on Raw Data Files}.
\newblock In {\em {Proceedings of 2012 ACM SIGMOD International Conference on
  Management of Data}}, pages 241--252, 2012.

\bibitem{christoph}
C.~Amb{\"u}hl, M.~Mastrolilli, and O.~Svensson.
\newblock {Inapproximability Results for Maximum Edge Biclique, Minimum Linear
  Arrangement, and Sparsest Cut}.
\newblock {\em {SIAM Journal on Computing}}, 40(2):567--596, 2011.

\bibitem{mip}
D.~Bertsimas and J.~N. Tsitsiklis.
\newblock {\em {Introduction to Linear Optimization}}.
\newblock Athena Scientific, 1997.

\bibitem{sdsq}
S.~Blanas, K.~Wu, S.~Byna, B.~Dong, and A.~Shoshani.
\newblock {Parallel Data Analysis Directly on Scientific File Formats}.
\newblock In {\em {Proceedings of 2014 ACM SIGMOD International Conference on
  Management of Data}}, pages 385--396, 2014.

\bibitem{offline-index-building}
N.~Bruno and S.~Chaudhuri.
\newblock {Automatic Physical Database Tuning: A Relaxation-Based Approach}.
\newblock In {\em {Proceedings of 2005 ACM SIGMOD International Conference on
  Management of Data}}, pages 227--238, 2005.

\bibitem{scanraw}
Y.~Cheng and F.~Rusu.
\newblock {Parallel In-Situ Data Processing with Speculative Loading}.
\newblock In {\em {Proceedings of 2014 ACM SIGMOD International Conference on
  Management of Data}}, pages 1287--1298, 2014.

\bibitem{chu:vert-part}
W.~Chu and I.~T. Ieong.
\newblock {A Transaction-Based Approach to Vertical Partitioning for Relational
  Database Systems}.
\newblock {\em IEEE Transactions on Software Engineering}, 19(8):804--812,
  1993.

\bibitem{psyu:mip}
D.~W. Cornell and P.~S. Yu.
\newblock {An Effective Approach to Vertical Partitioning for Physical Design
  of Relational Databases}.
\newblock {\em IEEE Transactions on Software Engineering}, 16(2):248--258,
  1990.

\bibitem{hyrise}
M.~Grund, J.~Kruger, H.~Plattner, A.~Zeier, P.~Cudre-Mauroux, and S.~Madden.
\newblock {HYRISE: A Main Memory Hybrid Storage Engine}.
\newblock {\em PVLDB}, 4(2):105--116, 2010.

\bibitem{hammer:vertical}
M.~Hammer and B.~Niamir.
\newblock {A Heuristic Approach to Attribute Partitioning}.
\newblock In {\em {Proceedings of 1979 ACM SIGMOD International Conference on
  Management of Data}}, pages 93--101, 1979.

\bibitem{data-morphing}
R.~Hankins and J.~Patel.
\newblock {Data Morphing: An Adaptive, Cache-Conscious Storage Technique}.
\newblock In {\em Proceedings of 2003 VLDB International Conference on Very
  Large Data Bases}, pages 417--428, 2003.

\bibitem{hoffer:mip}
J.~Hoffer.
\newblock {An Integer Programming Formulation of Computer Data Base Design
  Problems}.
\newblock {\em {Information Sciences}}, 11(1):29--48, 1976.

\bibitem{files-queries-results}
S.~Idreos, I.~Alagiannis, R.~Johnson, and A.~Ailamaki.
\newblock {Here are my Data Files. Here are my Queries. Where are my Results?}
\newblock In {\em {Proceedings of 2011 CIDR Conference on Innovative Database
  Research}}, pages 57--68, 2011.

\bibitem{tuple-reconstruction}
S.~Idreos, M.~L. Kersten, and S.~Manegold.
\newblock {Self-Organizing Tuple Reconstruction in Column-Stores}.
\newblock In {\em {Proceedings of 2009 ACM SIGMOD International Conference on
  Management of Data}}, pages 297--308, 2009.

\bibitem{data-vaults}
M.~Ivanova, M.~L. Kersten, and S.~Manegold.
\newblock {Data Vaults: A Symbiosis between Database Technology and Scientific
  File Repositories}.
\newblock In {\em Proceedings of 2012 SSDBM International Conference on
  Scientific and Statistical Database Management}, pages 485--494, 2012.

\bibitem{jindal:o2p}
A.~Jindal and J.~Dittrich.
\newblock {Relax and Let the Database Do the Partitioning Online}.
\newblock In {\em {Proceedings of 2011 BIRTE International Workshop on Enabling
  Real-Time Business Intelligence}}, pages 65--80, 2011.

\bibitem{vert-part:survey}
A.~Jindal, E.~Palatinus, V.~Pavlov, and J.~Dittrich.
\newblock {A Comparison of Knives for Bread Slicing}.
\newblock {\em PVLDB}, 6(6):361--372, 2013.

\bibitem{jindal:trojan}
A.~Jindal, J.~Quiane-Ruiz, and J.~Dittrich.
\newblock {Trojan Data Layouts: Right Shoes for a Running Elephant}.
\newblock In {\em {Proceedings of 2011 ACM SoCC Symposium on Cloud Computing}},
  2011.

\bibitem{raw}
M.~Karpathiotakis, M.~Branco, I.~Alagiannis, and A.~Ailamaki.
\newblock {Adaptive Query Processing on RAW Data}.
\newblock {\em PVLDB}, 7(12):1119--1130, 2014.

\bibitem{impala}
{M. Kornacker et al.}
\newblock {Impala: A Modern, Open-Source SQL Engine for Hadoop}.
\newblock In {\em {Proceedings of 2015 CIDR Conference on Innovative Database
  Research}}, 2015.

\bibitem{ailamaki:smdb:online-vert-part}
T.~Malik, X.~Wang, R.~Burns, D.~Dash, and A.~Ailamaki.
\newblock {Automated Physical Design in Database Caches}.
\newblock In {\em Proceedings of 2008 International Conference on Data
  Engineering SMDB Workshop}, pages 27--34, 2008.

\bibitem{ailamaki:ssdbm:online-vert-part}
T.~Malik, X.~Wang, D.~Dash, A.~Chaudhary, A.~Ailamaki, and R.~Burns.
\newblock {Adaptive Physical Design for Curated Archives}.
\newblock In {\em Proceedings of 2009 SSDBM International Conference on
  Scientific and Statistical Database Management}, pages 148--166, 2009.

\bibitem{instant-loading}
T.~M\"{u}hlbauer, W.~R\"{o}diger, R.~Seilbeck, A.~Reiser, A.~Kemper, and
  T.~Neumann.
\newblock {Instant Loading for Main Memory Databases}.
\newblock {\em PVLDB}, 6(14):1702--1713, 2013.

\bibitem{mysql:external-tables}
{MySQL 5.7 Manual}.
\newblock {The CSV Storage Engine}, 2013.

\bibitem{ptf:overview}
{N. M. Law et al.}
\newblock {The Palomar Transient Factory: System Overview, Performance and
  First Results}.
\newblock {\em CoRR}, abs/0906.5350, 2009.

\bibitem{vertical-part}
S.~Navathe, S.~Ceri, G.~Wiederhold, and J.~Dou.
\newblock {Vertical Partitioning Algorithms for Database Design}.
\newblock {\em {Transactions on Database Systems (TODS)}}, 9(4):680--710, 1984.

\bibitem{auto-part}
S.~Papadomanolakis and A.~Ailamaki.
\newblock {AutoPart: Automating Schema Design for Large Scientific Databases
  Using Data Partitioning}.
\newblock In {\em Proceedings of 2004 SSDBM International Conference on
  Scientific and Statistical Database Management}, pages 383--392, 2004.

\bibitem{minzheng}
M.~Z. Shieh, S.~C. Tsai, and M.~C. Yang.
\newblock {On the Inapproximability of Maximum Intersection Problems}.
\newblock {\em {Information Processing Letters}}, 112(19):723--727, 2012.

\bibitem{sdss:sqlLoader}
A.~Szalay, A.~Thakar, and J.~Gray.
\newblock {The sqlLoader Data-Loading Pipeline}.
\newblock {\em Computing in Science \& Engineering}, 10(1):38--48, 2008.

\bibitem{sinew}
D.~Tahara, T.~Diamond, and D.~Abadi.
\newblock {Sinew: A SQL System for Multi-Structured Data}.
\newblock In {\em {Proceedings of 2014 ACM SIGMOD International Conference on
  Management of Data}}, pages 815--826, 2014.

\bibitem{1000genomes:overview}
{The 1000 Genomes Project Consortium}.
\newblock {A Map of Human Genome Variation from Population-Scale Sequencing}.
\newblock {\em Nature}, 467(7319):1061--1073, 2010.

\bibitem{min-k-set-coverage}
S.~Vinterbo.
\newblock {Privacy: A Machine Learning View}.
\newblock {\em {Transactions on Knowledge and Data Engineering (TKDE)}},
  16(8):939--948, 2004.

\bibitem{oracle:external-tables}
A.~Witkowski, M.~Colgan, A.~Brumm, T.~Cruanes, and H.~Baer.
\newblock {Performant and Scalable Data Loading with Oracle Database 11g},
  2011.

\bibitem{lhc:overview}
A.~Wright and R.~Webb.
\newblock {The Large Hadron Collider}.
\newblock {\em Nature Insight}, 448(7151):269--312, 2007.

\bibitem{lsst}
{Z. Ivezic et al.}
\newblock {LSST: From Science Drivers to Reference Design and Anticipated Data
  Products}.
\newblock {\em CoRR}, abs/0805.2366, 2008.

\end{thebibliography}

\end{document}